\documentclass[traditabstract]{aa}
\usepackage{txfonts}
\usepackage{multirow}

\usepackage{multicol}

\usepackage{graphicx,longtable,lscape,natbib,amssymb,amssymb,amsmath}
\bibpunct{(}{)}{;}{a}{}{,} % to follow the A&A style

\usepackage{color}
\usepackage[citebordercolor={0 1 0}]{hyperref}

\newcommand{\be}{\begin{equation}}
\newcommand{\ee}{\end{equation}}
\newcommand{\ba}{\begin{eqnarray}}
\newcommand{\ea}{\end{eqnarray}}

\begin{document}
%\pdfoutput=1"

\title
{Impacts of different SNLS3 light-curve fitters on cosmological consequences of interacting dark energy models}

\author{Shuang Wang\inst{1}
          \and	
        Yazhou Hu\inst{1,2,3}
          \and
	    Miao Li\inst{1}
          \and	
        Nan Li\inst{1,2,3}
	  }

\institute{School of Physics and Astronomy, Sun Yat-Sen University, Guangzhou 510275, P. R. China\\
\email{wangshuang@mail.sysu.edu.cn,asiahu@itp.ac.cn, limiao9@mail.sysu.edu.cn, linan@itp.ac.cn}
           \and
           Kavli Institute for Theoretical Physics China,
Chinese Academy of Sciences, Beijing 100190, P. R. China\\
           \and
           State Key Laboratory of Theoretical Physics, Institute of Theoretical Physics,
           Chinese Academy of Sciences, Beijing 100190, P. R. China}

\date{\today}

\abstract{We explore the cosmological consequences of interacting dark energy (IDE) models using the SNLS3 supernova samples. In particular, we focus on the impacts of different SNLS3 light-curve fitters (LCF) (corresponding to ``SALT2'', ``SiFTO'', and ``Combined'' sample). Firstly, making use of the three SNLS3 data sets, as well as the Planck distance priors data and the galaxy clustering data, we constrain the parameter spaces of three IDE models. Then, we study the cosmic evolutions of Hubble parameter $H(z)$, deceleration diagram $q(z)$, statefinder hierarchy $S^{(1)}_3(z)$ and $S^{(1)}_4(z)$, and check whether or not these dark energy diagnosis can distinguish the differences among the results of different SNLS3 LCF. At last, we perform high redshift cosmic age test using three old high redshift objects (OHRO), and explore the fate of the Universe. We find that, the impacts of different SNLS3 LCF are rather small, and can not be distinguished by using $H(z)$, $q(z)$, $S^{(1)}_3(z)$, $S^{(1)}_4(z)$, and the age data of OHRO. In addition, we infer, from the current observations, how far we are from a cosmic doomsday in the worst case, and find that the ``Combined'' sample always gives the largest 2$\sigma$ lower limit of the time interval between ``big rip'' and today, while the results given by the ``SALT2'' and the ``SiFTO'' sample are close to each other. These conclusions are insensitive to a specific form of dark sector interaction.
Our method can be used to distinguish the differences among various cosmological observations.}


\titlerunning{Impacts of different SNLS3 LCF on cosmological consequences of IDE models}

\authorrunning{Wang el al.}

\maketitle
%----------------------------------------------------------------------

\section{Introduction}  \label{sec:intro}

Type Ia supernova is a sub-category of cataclysmic variable stars that results from the violent explosion of a white dwarf star in a binary system~\citep{Hillebrandt2000}.
Now it has become one of the most powerful probes to illuminate the mystery of cosmic acceleration \citep{Riess98,Perl99},
which may be due to an unknown energy component that can produce an anti-gravitational effect, i.e., dark energy (DE), or a modification of general relativity, i.e., modified gravity (MG).
\footnote{For recent reviews, see \cite{Frieman08,Caldwell09,Uzan10,Wang10,LLWW11,LLWW13,Weinberg13}.}
Along with the rapid progress of supernova (SN) cosmology, several high quality SN datasets were released in recent years,
such as ``SNLS'' \cite{Astier06}, ``Union'' \cite{Kowalski08}, ``Constitution'' \cite{Hicken09}, ``SDSS'' \cite{Kessler09},
``Union2'' \cite{Amanullah10}, ``Union2.1'' \cite{Suzuki12}, and ``Pan-STARRS1'' \cite{Scolnic14}.
In 2010, the Supernova Legacy Survey (SNLS) group released their three years data \cite{Guy10}.
Soon after, combining these SN samples with various low-$z$ to mid-$z$ samples
and making use of three different light-curve fitters (LCF)~\footnote{A LCF is a method to model the light curves of SN; by using it one can estimate the distance information of each SN.},
Conley et al. presented three SNLS3 data sets \cite{Conley11}:
``SALT2'', which consists of 473 SN;
``SiFTO'', which consists of 468 SN;
and ``Combined'', which consists of 472 SN.
It should be mentioned that, in the cosmology fits, the SNLS group treated two important quantities,
stretch-luminosity parameter $\alpha$ and color-luminosity parameter $\beta$ of SNe Ia, as free model parameters.
Notice that $\alpha$ and $\beta$ are parameters for the luminosity standardization.
In addition, the intrinsic scatter $\sigma_{int}$ are fixed to ensure that $\chi^2/dof=1$.

A most critical challenge of SN cosmology is the control of the systematic uncertainties of type Ia supernovae (SNe Ia).
One of the most important factors is the potential SN evolution.
For examples, previous studies on the Union2.1 \cite{Mohlabeng13} and the Pan-STARRS1 data sets \cite{Scolnic14}
all indicated that $\beta$ should evolve along with redshift $z$.
Besides, it was found that the intrinsic scatter $\sigma_{\rm int}$ has the hint of redshift-dependence that will significantly affect the results of cosmology fits~\citep{Marrinerl11}.
In addition to the SN evolution,
another important factor is the choice of LCF~\citep{Kessler09}.
For instance, it has been proved that even for the same SN samples,
using ``MLCS2k2'' \citep{Jha07} and ``SALT2'' \citep{Guy07} LCF will lead to
completely different fitting results for various cosmological models \citep{Sollerman09,Sanchez09,Pigozzo11,Smale11,Bengochea11,Bengochea14}.

One of the present authors has also done a series of research works to study the systematic uncertainties of SNe Ia.
By using the SNLS3 dataset, we found that $\alpha$ is still consistent with a constant,
but $\beta$ evolves along with $z$ at very high confidence level (CL) \cite{WangWang13}.
Soon after, we showed that this result has significant effects on the parameter estimation of standard cosmology~\citep{WLZ14},
and the introduction of a time-varying $\beta$ can reduce the tension between SNe Ia and other cosmological observations.
Moreover, we proved that our conclusion holds true for various DE and MG models \citep{WWGZ14,WWZ14,Wang15}.
Besides, the evolution of $\alpha$ has also been found for the JLA sampls \citep{Li16}.
In addition, by using three different SNLS3 LCF, including ``SALT2'', ``SiFTO'' and ``Combined'',
we briefly discussed the effects of different LCF on the parameter estimation of the $\Lambda$-cold-dark-matter ($\Lambda$CDM) model \cite{WangWang13} and the holographic dark energy (HDE) model \citep{Wang15}.
It must be emphasized that in these two papers,
only the fitting results given by different SNLS3 LCF are shown, and the corresponding cosmological consequences are not discussed.
Therefore, the impacts of different LCF are not studied in details in our previous works.
The main scientific objective of the current work
is presenting a comprehensive and systematic investigation on the impacts of different SNLS3 LCF.
To do this, both the cosmology-fit results and the cosmological consequence results are shown in this work.

As mentioned above, the effects of different SNLS3 LCF on the $\Lambda$CDM model has been briefly discussed in~\citep{WangWang13}.
To obtain new scientific results, new elements need to be taken into account.
Since the interaction between different components widely exist in nature,
and the introduction of a interaction between DE and CDM can provide an intriguing mechanism to solve the ``cosmic coincidence problem''~\citep{Guo07,LLWZ09,Li09,He10} and alleviate the ``cosmic age problem''~\citep{WZ08, WLL10, CuiZhang2010, Duran2011},
it is interesting to study the effect of different LCF on interacting dark energy (IDE) model.
Here we adopt the $w$-cold-dark-matter ($w$CDM) model with a direct non-gravitational interaction between dark sectors.
It must be emphasized that,
if our conclusions are dependent on the specific form of dark sector interaction,
these conclusions will not be reliable at all.
So in this work, three kinds of interaction terms are taken into account
to ensure that our study is insensitive to a specific interaction form.
In addition, to make a comparison, we also consider the case of $w$CDM model without interaction term.

According to the previous studies on the potential SN evolution,
we adopt a constant $\alpha$ and a linear $\beta(z) = \beta_{0} + \beta_{1} z$ in this work.
Making use of the three SNLS3 data sets,
as well as the latest Planck distance prior data \citep{Wang13},
the galaxy clustering (GC) data extracted from
Sloan digital sky survey (SDSS) data release 7 (DR7) \citep{HWC14} and data release 9 (DR9) \citep{Wang14},
we constrain the parameter spaces of the $w$CDM model and the three IDE models,
and investigate the impacts of different SNLS3 LCF on the cosmology fits.
Moreover, based on the fitting results,
we study the possibility of distinguishing the impacts of different LCF by using various DE diagnosis tools and cosmic age data,
and discuss the possible fate of the Universe.

We present our method in Section~\ref{sec:method}, our results in Section~\ref{sec:results}, and
summarize and conclude in Section~\ref{sec:conclusion}.

\section{Methodology}
\label{sec:method}

In this section, firstly we review the theoretical framework of the IDE models,
then we briefly describe the observational data used in the present work,
and finally we introduce the background knowledge about DE diagnosis and cosmic age.

\subsection{Theoretical Models}
\label{subsec:model}

This work is a sequel to the previous studies of our group,
and thus we study the same IDE models considered in~\citep{WWGZ14}.
It must be mentioned that, here we consider a flat Universe and treat the present fractional density of radiation as a model parameter, which is a little different from the case of ~\citep{WWGZ14}.
In a flat Universe\footnote{The assumption of flatness is motivated by the inflation scenario. For a detailed discussion of the effects of spatial curvature, see \citep{Clarkson07}}, the Friedmann equation is
\be\label{F.e.}
    3M_{pl}^{2}H^{2}=\rho_{c}+\rho_{de}+\rho_{r}+\rho_{b},
\ee
where $H \equiv \dot{a}/a$ is the Hubble parameter,
$a=(1+z)^{-1}$ is the scale factor of the Universe (we take today's scale factor $a_0=1$),
the dot denotes the derivative with respect to cosmic time $t$,
$M^2_p = (8\pi G)^{-1}$ is the reduced Planck mass, $G$ is Newtonian gravitational constant,
$\rho_{c}$, $\rho_{de}$, $\rho_{r}$ and $\rho_{b}$
are the energy densities of CDM, DE, radiation and baryon, respectively.
The reduced Hubble parameter $E(z)\equiv H(z)/H_{0}$ satisfies
\be\label{E}
E^{2}=\Omega_{c0}\frac{\rho_{c}}{\rho_{c0}}+\Omega_{de0}\frac{\rho_{de}}{\rho_{de0}}
+\Omega_{r0}\frac{\rho_{r}}{\rho_{r0}}+\Omega_{b0}\frac{\rho_{b}}{\rho_{b0}},
\ee
where $\Omega_{c0}$, $\Omega_{de0}$, $\Omega_{r0}$ and $\Omega_{b0}$
are the present fractional densities of CDM, DE, radiation and baryon, respectively.
As mentioned above, we treat $\Omega_{r0}$ as a model parameter in this work.
In addition, $\rho_{r}=\rho_{r0}(1+z)^{4}$, $\rho_{b}=\rho_{b0}(1+z)^{3}$.
Since $\Omega_{de0}=1-\Omega_{c0}-\Omega_{b0}-\Omega_{r0}$, $\Omega_{de0}$ is not an independent parameter.

In an IDE scenario, the energy conservation equations of CDM and DE satisfy
\ba
\label{eq:CEt1}&& \dot \rho_c+3H\rho_c=Q,\ \ \\
\label{eq:CEt2}&& \dot \rho_{de}+3H(\rho_{de}+p_{de})=-Q,
\ea
where $p_{de} = w\rho_{de}$ is the pressure of DE,
$Q$ is the interaction term, which describes the energy transfer between CDM and DE.
So far, the microscopic origin of interaction between dark sectors is still a puzzle.
To study the issue of interaction, one needs to write down the possible forms of $Q$ by hand.
In this work we consider the following three cases:
\ba
    &&Q_{1}=3\gamma H\rho_{c},\ \ \\
    &&Q_{2}=3\gamma H\rho_{de},\ \ \\
    &&Q_{3}=3\gamma H\frac{\rho_{c}\rho_{de}}{\rho_{c}+\rho_{de}},\ \
\ea
where $\gamma$ is a dimensionless parameter describing the strength of interaction,
$\gamma > 0$ means that energy transfers from DE to CDM,
$\gamma < 0$ implies that energy transfers from CDM to DE, and $\gamma=0$ denotes the case without dark sector interaction, i.e. the $w$CDM model.
Notice that these three kinds of interaction form have been widely studied in the literature ~\citep{Guo07,Li09,He10,LZ14}.
For simplicity, hereafter we call them I$w$CDM1 model,  I$w$CDM2 model, and I$w$CDM3 model, respectively.

For the $w$CDM model, we have
\ba\label{E:Q0}
E(z)=\Big (\Omega_{r0}(1+z)^{4}+(\Omega_{b0}+\Omega_{c0})(1+z)^{3}+\Omega_{de0}(1+z)^{3(1+w)}\Big )^{1/2}.
\ea
For the I$w$CDM1 model, the solutions of Eqs. (\ref{eq:CEt1}) and (\ref{eq:CEt2}) are
\be
\label{rho:c1}
\rho_{c}=\rho_{c0}(1+z)^{3(1-\gamma)},
\ee
\be
\label{rho:de1}
\rho_{de}=\frac{\gamma\rho_{c0}}{w+\gamma}\big((1+z)^{3(1+w)}-(1+z)^{3(1-\gamma)}\big) +\rho_{de0}(1+z)^{3(1+w)}.
\ee
Then we have
\ba\label{E:Q1}
E(z)=\Big (\Omega_{r0}(1+z)^{4}+\Omega_{b0}(1+z)^{3}+\Omega_{de0}(1+z)^{3(1+w)} \nonumber\\
 +\Omega_{c0}\big (\frac{\gamma}{w+\gamma}(1+z)^{3(1+w)}+\frac{w}{w+\gamma}(1+z)^{3(1-\gamma)}\big )\Big )^{1/2}.
\ea
\\
For the I$w$CDM2 model, the solutions of Eqs. (\ref{eq:CEt1}) and (\ref{eq:CEt2}) are
\be
\label{rho:de2}
\rho_{de}=\rho_{de0}(1+z)^{3(1+w+\gamma)},
\ee
\be
\label{rho:c2}
\rho_{c}=\rho_{c0}(1+z)^{3}+\frac{\gamma\rho_{de0}}{w+\gamma}(1+z)^{3}-\frac{\gamma\rho_{de0}}{w+\gamma}(1+z)^{3(1+w+\gamma)}.
\ee
Then we get
\ba\label{E:Q2}
E(z)=\Big (\Omega_{r0}(1+z)^{4}+(\Omega_{c0}+\Omega_{b0})(1+z)^{3}  \nonumber\\
+\Omega_{de0}\big (\frac{\gamma}{w+\gamma}(1+z)^{3}+ \frac{w}{w+\gamma}(1+z)^{3(1+w+\gamma)} \big )\Big )^{1/2}.
\ea
\\
For the I$w$CDM3 model, Eqs. (\ref{eq:CEt1}) and (\ref{eq:CEt2}) still have analytical solutions
\be
\label{rho:c3}
\rho_{c}=\rho_{c0}(1+z)^{3}\big(\frac{\rho_{c0}}{\rho_{c0}+\rho_{de0}}  +\frac{\rho_{de0}}{\rho_{c0}+\rho_{de0}}(1+z)^{3(w+\gamma)}\big)^{-\frac{\gamma}{w+\gamma}},
\ee
\ba
\label{rho:de3}
\rho_{de}=\rho_{de0}(1+z)^{3(1+w+\gamma)}\big(\frac{\rho_{c0}}{\rho_{c0}+\rho_{de0}} \nonumber\\ +\frac{\rho_{de0}}{\rho_{c0}+\rho_{de0}}(1+z)^{3(w+\gamma)}\big)^{-\frac{\gamma}{w+\gamma}}.
\ea
Then we obtain
\ba\label{E:Q3}
E(z)=\big(\Omega_{r0}(1+z)^{4}+\Omega_{b0}(1+z)^{3}+\Omega_{c0}C(z)(1+z)^{3}  \nonumber\\
+\Omega_{de0}C(z)(1+z)^{3(1+w+\gamma)}\big)^{1/2}.
\ea
where
\be\label{Cz}
C(z)=\big(\frac{\Omega_{c0}}{\Omega_{c0}+\Omega_{de0}}+\frac{\Omega_{de0}}{\Omega_{c0}+\Omega_{de0}}(1+z)^{3(w+\gamma)}\big)^{-\frac{\gamma}{w+\gamma}}.
\ee
For each model, the expression of $E(z)$ will be used to calculate the observational quantities appearing in the next subsection.

\subsection{Observational Data}

In this subsection, we describe the observational data.
Here we use the SNe Ia, CMB, and GC data, which is same with our previous paper~\citep{WLZ14}.
It should be emphasized that, in this work we use all the three SNLS3 data sets (i.e. ``SALT2'', ``SiFTO'', and ``Combined'' sample), while only the ``combined'' data are used in~\citep{WLZ14}. in addition, the GC data are also updated in this work, compared to that used in~\citep{WLZ14}.

\subsubsection{SNe Ia Data}

As mentioned above, we use all the three SNLS3 data sets.
In the following, we briefly introduce how to include these three data sets into the $\chi^2$ analysis.

Adopting a constant $\alpha$ and a linear $\beta(z) = \beta_{0} + \beta_{1} z$,
the predicted magnitude of an SN becomes
\be
m_{\rm mod}=5 \log_{10}{\cal D}_L(z) - \alpha (s-1) +\beta(z) {\cal C} + {\cal M}.
\ee
The luminosity distance ${\cal D}_L(z)$ is defined as
\be
{\cal D}_L(z)\equiv H_0 (1+z_{\rm hel}) r(z),
\ee
where $z$ and $z_{\rm hel}$ are the CMB restframe and heliocentric redshifts of SN,
and
\be
r(z)=H_0^{-1}\int_0^z\frac{dz'}{E(z')}.
\ee
Here $s$ and ${\cal C}$ are stretch measure and color measure for the SN light curve,
${\cal M}$ is a parameter representing some combination of SN absolute magnitude $M$ and Hubble constant $H_0$.

For a set of $N$ SNe with correlated errors, the $\chi^2$ function is given by
\be
\label{eq:chi2_SN}
\chi^2_{SN}=\Delta \mbox{\bf m}^T \cdot \mbox{\bf C}^{-1} \cdot \Delta\mbox{\bf m},
\ee
where $\Delta m \equiv m_B-m_{\rm mod}$ is a vector with $N$ components,
and $m_B$ is the rest-frame peak B-band magnitude of the SN.
The total covariance matrix $\mbox{\bf C}$ can be written as \cite{Conley11}
\be
\mbox{\bf C}=\mbox{\bf D}_{\rm stat}+\mbox{\bf C}_{\rm stat}+\mbox{\bf C}_{\rm sys}.
\ee
Here $\mbox{\bf D}_{\rm stat}$ denotes the diagonal part of the statistical uncertainty,
$\mbox{\bf C}_{\rm stat}$ and $\mbox{\bf C}_{\rm sys}$ denote the statistical and systematic covariance matrices, respectively.
For the details of constructing the covariance matrix $\mbox{\bf C}$, see \cite{Conley11}.

It must be emphasized that,
in order to include host-galaxy information in the cosmological fits,
Conley et al. split the SNLS3 sample based on host-galaxy stellar mass at $10^{10} M_{\odot}$,
and made ${\cal M}$ to be different for the two samples \cite{Conley11}.
So there are two values of ${\cal M}$ (i.e. ${\cal M}_1$ and ${\cal M}_2$) for the SNLS3 data.
Moreover, Conley et al. removed ${\cal M}_1$ and ${\cal M}_2$ from cosmology-fits by analytically marginalizing over them
(for more details, see the appendix C of \cite{Conley11}).
In the present work, we just follow the recipe of \cite{Conley11}, and do not treat ${\cal M}$ as model parameter.

\subsubsection{CMB Data}

For CMB data, we use the distance priors data extracted from Planck first data release \citep{Wang13}
\footnote{It should be mentioned that, in this paper we just use the purely geometric measurements of CMB, i.e. the distance prior data.
There are some other methods of using CMB data.
For examples, the observed position of the first peak of the CMB anisotropies spectrum can be used to perform cosmology-fits \citep{Carneiro08,Pigozzo11}.
In addition, the CMB full data can also be used to constrain cosmological models via the global fit technique.
To make a comparison,
we constrain the parameter spaces of I$w$CDM1 model by using these three methods of using CMB data.
We find that the differences among the fitting results given by different method are very small
(For example, the differences on DE EoS $w$ are only the order of $1\%$).
In other words, Our results are insensitive to the method of using CMB data.
This conclusion also holds true for other DE models (such as HDE model \citep{LWLZ13}),
showing that the CMB data cannot put strict constraints on the properties of DE.
Since the main purpose of using CMB data is to put strict constraints on $\Omega_{c0}$ and $\Omega_{b0}$,
we think that the use of the Planck distance prior is sufficient enough for our work.
}.
CMB give us the comoving distance to the photon-decoupling surface $r(z_*)$\footnote{In this work, we adopt the result of $z_*$ given in \citep{Hu96}.},
and the comoving sound horizon at photon-decoupling epoch $r_s(z_*)$.
Wang and Mukherjee showed that the CMB shift parameters \citep{Wang07}
\ba
l_a &\equiv &\pi r(z_*)/r_s(z_*), \nonumber\\
R &\equiv &\sqrt{\Omega_m H_0^2} \,r(z_*)/c,
\ea
together with $\omega_b\equiv \Omega_b h^2$, provide an efficient summary
of CMB data as far as dark energy constraints go.
The comoving sound horizon is given by \citep{Wang13}
\be
\label{eq:rs}
r_s(z)  =  cH_0^{-1} \int_0^{a} \frac{da'}{\sqrt{ 3(1+ \overline{R_b}\,a')\, {a'}^4 E^2(z')}},
\ee
where $a$ is the scale factor of the Universe,
$\overline{R_b}=31500\Omega_bh^2(T_{\rm cmb}/2.7\,{\rm K})^{-4}$,
and $T_{\rm cmb}=2.7255\,{\rm K}$.

Using the Planck+lensing+WP data,
the mean values and 1$\sigma$ errors of $\{ l_a, R, \omega_b\}$ are obtained \citep{Wang13},
\ba
&&\langle l_a \rangle = 301.57, \sigma(l_a)=0.18, \nonumber\\
&&\langle R \rangle = 1.7407,  \sigma(R)=0.0094, \nonumber\\
&& \langle \omega_b \rangle = 0.02228, \sigma(\omega_b)=0.00030.
\label{eq:CMB_mean_planck}
\ea
Defining $p_1=l_a(z_*)$, $p_2=R(z_*)$, and $p_3=\omega_b$,
the normalized covariance matrix $\mbox{NormCov}_{CMB}(p_i,p_j)$ can be written as \citep{Wang13}
\be
\left(
\begin{array}{ccc}
   1.0000  &    0.5250  &   -0.4235    \\
  0.5250  &     1.0000  &   -0.6925    \\
 -0.4235  &   -0.6925  &     1.0000    \\
\end{array}
\right).
\label{eq:normcov_planck}
\ee
Then, the covariance matrix for $(l_a, R, \omega_b)$ is given by
\be
\mbox{Cov}_{CMB}(p_i,p_j)=\sigma(p_i)\, \sigma(p_j) \,\mbox{NormCov}_{CMB}(p_i,p_j),
\label{eq:CMB_cov}
\ee
where $i,j=1,2,3$.
The CMB data are included in our analysis by adding
the following term to the $\chi^2$ function:
\be
\label{eq:chi2CMB}
\chi^2_{CMB}=\Delta p_i \left[ \mbox{Cov}^{-1}_{CMB}(p_i,p_j)\right]
\Delta p_j,
\hskip .2cm
\Delta p_i= p_i - p_i^{data},
\ee
where $p_i^{data}$ are the mean values from Eq. (\ref{eq:CMB_mean_planck}).

\subsubsection{GC Data}

To improve the cosmological constraints, we also use the GC data extracted from SDSS samples.
In \citep{CW13},
Chuang and Wang measured the Hubble parameter $H(z)$ and the angular-diameter distance $D_A(z)$ separately
\footnote{Angular-diameter distance $D_A(z)= c H_0^{-1} r(z)/(1+z)$, where $c$ is the speed of light.}
by scaling the model galaxy two-point correlation function to match the observed galaxy two-point correlation function.
Since the scaling is measured by marginalizing over the shape of the model correlation function,
the measured $H(z)$ and $D_A(z)$ are model-independent and can be used to constrain any cosmological model \citep{CW13}.

It should be mentioned that, compared with using $H(z)$ and $D_A(z)$,
using $H(z)r_s(z_d)/c$ and $D_A(z)/r_s(z_d)$
\footnote{$r_s(z_d)$ is the sound horizon at the drag epoch, where $r_s(z)$ is given by Eq. (\ref{eq:rs}, $z_d$ is given in \citep{Eisenstein98}.}
can give better constraints on various cosmological models.
In \citep{HWC14}, using the two-dimensional matter power spectrum of SDSS DR7 samples,
Hemantha, Wang, and Chuang got
\ba
H(z=0.35)r_s(z_d)/c&=&0.0431  \pm  0.0018,  \nonumber \\
D_A(z=0.35)/r_s(z_d)&=& 6.48  \pm  0.25.
\label{eq:CW2}
\ea
In a similar work \citep{Wang14}, using the anisotropic two-dimensional galaxy correlation function of SDSS DR9 samples,
Wang obtained
\ba
H(z=0.57)r_s(z_d)/c&=&0.0444	\pm  0.0019,  \nonumber \\
D_A(z=0.57)/r_s(z_d)&=& 9.01	\pm  0.23.
\label{eq:C13}
\ea
GC data are included in our analysis by adding $\chi^2_{GC}=\chi^2_{GC1}+\chi^2_{GC2}$,
with $z_{GC1}=0.35$ and $z_{GC2}=0.57$, to the $\chi^2$ of a given model.
Note that
\be
\label{eq:chi2bao}
\chi^2_{GCi}=\Delta q_i \left[ {\rm C}^{-1}_{GCi}(q_i,q_j)\right]
\Delta q_j,
\hskip .2cm
\Delta q_i= q_i - q_i^{data},
\ee
where $q_1=H(z_{GCi})r_s(z_d)/c$, $q_2=D_A(z_{GCi})/r_s(z_d)$, and $i=1,2$.
Based on Refs. \citep{HWC14} and \citep{Wang14}, we have
\begin{equation}
 {\rm C}_{GC1}=\left(
  \begin{array}{cc}
    0.00000324 & -0.00010728 \\
    -0.00010728 & 0.0625 \\
  \end{array}
\right),
\end{equation}
\begin{equation}
 {\rm C}_{GC2}=\left(
  \begin{array}{cc}
    0.00000361 & 0.0000176111 \\
    0.0000176111 & 0.0529 \\
  \end{array}
\right).
\end{equation}

\subsubsection{Total $\chi^2$ Function}

Now the total $\chi^2$ function is
\be
\chi^2=\chi^2_{SN}+\chi^2_{CMB}+\chi^2_{GC}.
\ee
We perform an MCMC likelihood analysis using the ``CosmoMC'' package  \citep{Lewis02}.

\subsection{Dark Energy Diagnosis and Cosmic Age}

Since previous works pointed out that the differences among the cosmological results given by the three different SNLS3 LCF are rather small, we need more tools to distinguish the effects of different LCF.
In the present work, we use the Hubble parameter $H(z)$, the deceleration parameter $q(z)$, the statefinder hierarchy $S^{(1)}_3(z)$ and $S^{(1)}_4(z)$~\citep{ArabSahni11,ZCZ14}, and the cosmic age $t(z)$ as our diagnosis tools to distinguish the effects of different LCF.
This subsection consists of two parts.
Firstly, we introduce the tools of DE diagnosis,
including the Hubble parameter $H(z)$, the deceleration parameter $q(z)$,
and the statefinder hierarchy $\{S^{(1)}_3, S^{(1)}_4\}$.
Then, we discuss the issues about cosmic age,
including the high-redshift cosmic age test and the fate of the Universe.

Let us introduce the DE diagnosis tools first.
The scale factor of the Universe $a$ can be Taylor expanded around today's cosmic age $t_0$ as follows:
\be
a(t)=1+\sum\limits_{\emph{n}=1}^{\infty}\frac{A_{\emph{n}}}{n!}[H_0(t-t_0)]^n,
\ee
where
\be
A_{\emph{n}}=\frac{a(t)^{(n)}}{a(t)H^n},~~n\in N,
\ee
with $a(t)^{(n)}=d^na(t)/dt^n$.
The Hubble parameter $H(z)$ contain the information of the first derivative of $a(t)$.
Based on the Baryon Acoustic Oscillations (BAO) measurements from the SDSS data release 9 and data release 11,
Samushia et al. gave $H_{0.57}\equiv H(z=0.57)=92.4\pm4.5 {\rm km/s/Mpc}$ \cite{Samushia13},
while Delubac et al. obtained $H_{2.34}\equiv H(z=2.34)=222\pm7 {\rm km/s/Mpc}$ \cite{Delubac14}.
These two H(z) data points will be used to compare the theoretical predictions of the $w$CDM model and the three IDE models.
In addition, the deceleration parameter $q$ is given by
\be
q=-A_2=-\frac{\ddot{a}}{aH^{2}},
\ee
which contains the information of the second derivatives of $a(t)$.
For the $\Lambda$CDM model,
$A_{2}|_{\Lambda \rm{CDM}}=1-\frac{3}{2}\Omega_{m}$, $A_{3}|_{\Lambda \rm{CDM}}=1$, $A_{4}|_{\Lambda \rm{CDM}}=1-\frac{3^2}{2}\Omega_{m}$.
The statefinder hierarchy, $S_{\emph{n}}$, is defined as \cite{ArabSahni11}:
\ba
&S_{2}=A_{2}+\frac{3}{2}\Omega_{\rm m},\\
&S_{3}=A_{3},\\
&S_{4}=A_{4}+\frac{3^2}{2}\Omega_{\rm m},
\ea
The reason for this redefinition is to peg the statefinder at unity for $\Lambda$CDM during the cosmic expansion,
\be
S_{\emph{n}}|_{\Lambda \rm{CDM}}=1.
\ee
This equation defines a series of null diagnostics for $\Lambda$CDM when $n\geq3$.
By using this diagnostic, we can easily distinguish the $\Lambda$CDM model from other DE models.
Because of $\Omega_{m}|_{\Lambda \rm{CDM}}=\frac{2}{3}(1+q)$, when $n\geq3$, statefinder hierarchy can be rewritten as:
\ba
&S^{(1)}_{3}=A_{3},\\
&S^{(1)}_{4}=A_{4}+3(1+q),
\ea
where the superscript $(1)$ is to discriminate between $S^{(1)}_{\emph{n}}$ and $S_{\emph{n}}$.
In this paper, we use the statefinder hierarchy $S^{(1)}_3(z)$ and $S^{(1)}_4(z)$ as our diagnosis tools to distinguish the effects of different LCF.

Now, let us turn to the issues about cosmic age.
The age of the Universe at redshift $z$ is given by
\be
t(z) =\int_z^\infty\frac{d\tilde{z}}{(1+\tilde{z})H(\tilde{z})}.
\ee
In history, the cosmic age problem played an important role in the cosmology \cite{Alcaniz99,LLLW10,Bengaly14,Liu14}).
Obviously, the Universe cannot be younger than its constituents.
In other words, the age of the universe at any high redshift $z$ cannot be younger than its constituents at the same redshift.
There are some old high redshift objects (OHRO) considered extensively in the literature.
For instance,
the 3.5Gyr old galaxy LBDS 53W091 at redshift $z = 1.55$ \cite{Dunlop96},
the 4.0Gyr old galaxy LBDS 53W069 at redshift $z = 1.43$ \cite{Dunlop99},
and the old quasar APM 08279+5255, whose age is estimated to be $2.0$ Gyr, at redshift $z = 3.91$ \cite{Hasinger02}.
In the literature, the age data of these three OHRO
(i.e. $t_{1.43} \equiv t(z=1.43)=4.0 Gyr$, $t_{1.55} \equiv t(z=1.55)=3.5 Gyr$ and $t_{3.91} \equiv t(z=3.91)=2.0 Gyr$)
have been extensively used to test various cosmological models (see e.g. \cite{Alcaniz03,Wei07,WZ08,WLL10,Yan14}).
In the present work, we will use these three age data to test the $w$CDM model and the three IDE models.

Another interesting topic is the fate of the Universe.
The future of the Universe depends on the property of DE.
If the Universe is dominated by a quintessence \cite{Caldwell98,Zlatev99} or a cosmological constant,
the expansion of the Universe will continue forever.
If the Universe is dominated by a phantom \cite{Caldwell02,Caldwell03},
eventually the repulsive gravity of DE will become large enough to tear apart all the structures,
and the Universe will finally encounter a doomsday, i.e. the so-called ``big rip'' (BR).
Setting $x=-ln(1+z)$, we can get the time interval between a BR and today
\be
\label{eq:BR}
t_{BR}-t_0 =\int_0^\infty\frac{dx}{H(x)},
\ee
where $t_{BR}$ denotes the time of BR.
It is obvious that, for a Universe dominated by a quintessence or a cosmological constant, this integration is infinity;
and for a Universe dominated by a phantom, this integration is convergence.
We would like to infer, from the current observational data, how far we are from a cosmic doomsday in the worst case.
So in this work we calculate the 2$\sigma$ lower limits of $t_{BR}-t_0$ for all the four DE models.

\section{Result}
\label{sec:results}

In this section, firstly we show the cosmology-fit results of the $w$CDM model
given by various SNLS3 samples without and with systematic uncertainties, next
we present the fitting results of the three IDE models,
then we show the cosmic evolutions of Hubble parameter $H(z)$,
deceleration parameter $q(z)$, statefinder hierarchy $S^{(1)}_3(z)$ and $S^{(1)}_4(z)$ according to the fitting results of the IDE models,
finally we perform the high-redshift cosmic age test and discuss the fate of the Universe.
Since both three SNLS3 datasets and a time-varying $\beta$ are considered at the same time,
all the results of this work are new compared with the previous studies.

\subsection{Cosmology Fits}

\begin{table*}

\caption{Fitting results for the $w$CDM model given by various SNLS3 samples without and with systematic uncertainties.
Both the best-fit values and the 1$\sigma$ errors of $\Omega_{m0}$ and $w$ are listed.
``Stat Only'' and  ``Stat Plus Sys'' represent the SN data only including
statistical uncertainties and the SN data including both statistical and
systematic uncertainties, respectively. Moreover, for comparison, both the cases of constant $\beta$ and linear $\beta$ are taken into account.}

\label{tab:SN}
\center
\begin{tabular}{ccccc}
\hline\hline
 & &Combined & SALT2 & SiFTO \\
\hline
\multirow{2}{*}{Constant $\beta$ (Stat Only)}
& $\Omega_{m0}$& $0.188^{+0.079}_{-0.059}$ & $0.216^{+0.069}_{-0.047}$  &  $0.178^{+0.105}_{-0.141}$  \\
\cline{2-5}
& $w$ & $-0.88^{+0.14}_{-0.11}$ & $-0.93^{+0.14}_{-0.12}$  &  $-0.84^{+0.19}_{-0.22}$ \\
\hline
\multirow{2}{*}{Constant $\beta$ (Stat Plus Sys)}  & $\Omega_{m0}$& $0.167^{+0.084}_{-0.070}$ & $0.236^{+0.083}_{-0.060}$  & $0.220^{+0.083}_{-0.061}$   \\
\cline{2-5}
& $w$ & $-0.88^{+0.18}_{-0.11}$ & $-0.99^{+0.19}_{-0.16}$  & $-1.01^{+0.19}_{-0.15}$  \\
\hline
\multirow{2}{*}{linear $\beta$ (Stat Plus Sys)}
 & $\Omega_{m0}$& $0.171^{+0.085}_{-0.070}$ & $0.186^{+0.095}_{-0.13}$  & $0.21^{+0.11}_{-0.11}$   \\
\cline{2-5}
& $w$ & $-0.88^{+0.18}_{-0.12}$ & $-0.856^{+0.22}_{-0.096}$  & $-0.96^{+0.26}_{-0.11}$  \\
\hline

\end{tabular}
\end{table*}

In table \ref{tab:SN}, we present the fitting results of the $w$CDM model given by
various SNLS3 samples without and with systematic uncertainties.
The first row of table \ref{tab:SN} shows the fitting results for the case of only
considering constant $\beta$ and statistical uncertainties. From this row we find that the
best-fit values of $\Omega_{m0}$ and $w$ given by the Combined sample are in
between the best-fit results given by the SALT2 sample and by the SiFTO sample.
This result is consistent with the result of \cite{Conley11}. The second row of table \ref{tab:SN}
shows the fitting results for the case of considering constant $\beta$ and
statistical+systematic uncertainties. From this row we can see that,
once the systematic uncertainties of SNLS3 samples are taken into account, the
best-fit values of $\Omega_{m0}$ and $w$ given by the Combined sample are not in between
the best-fit results given by the SALT2 sample and by the SiFTO sample any more.
Therefore, the reason causing this strange phenomenon is due to the systematic uncertainties of SNLS3 samples.
To further study this issue, in the third row of table \ref{tab:SN},
we present the fitting results for the case of considering linear $\beta$ and statistical+systematic uncertainties.
We find that, after considering the evolution of $\beta$, the best-fit value of $w$ given by the Combined sample is
in between the results given by the SALT2 sample and by the SiFTO sample, while the differences among the best-fit
values of $\Omega_{m0}$ given by the three SNLS3 samples are effectively reduced.
These results imply the importance of considering $\beta$'s evolution in the cosmology-fits.
Therefore, from now on we only consider the case of linear $\beta$.

\begin{table*}\scriptsize
\caption{Fitting results for the $w$CDM model and the three IDE models,
where both the best-fit values and the 1$\sigma$ errors of various parameters are listed.
``Combined'', ``SALT2'' and ``SiFTO'' represent the SN(Combined)+CMB+GC, the SN(SALT2)+CMB+GC and the SN(SiFTO)+CMB+GC data, respectively.}

\label{tab:combined_results}

\begin{tabular}{cp{1.0cm}p{1.0cm}p{1.0cm}cp{1.0cm}p{1.1cm}p{1.0cm}cp{1.0cm}p{1.0cm}p{1.0cm}cp{1.0cm}p{1.0cm}p{1.0cm}}
\hline\hline &\multicolumn{3}{c}{$w$CDM}&&\multicolumn{3}{c}{I$w$CDM1}&&\multicolumn{3}{c}{I$w$CDM2}&&\multicolumn{3}{c}{I$w$CDM3} \\
  \cline{2-4}\cline{6-8}\cline{10-12}\cline{14-16}
Parm  & Combined & SALT2 & SiFTO && Combined & SALT2 & SiFTO & & Combined & SALT2 & SiFTO & & Combined & SALT2 & SiFTO  \\ \hline
  $\alpha$         & $1.417^{+0.068}_{-0.071}$ %
                   & $1.576^{+0.135}_{-0.125}$  %
                   & $1.360^{+0.049}_{-0.046}$ &  %
                   & $1.406^{+0.080}_{-0.059}$
                   & $1.597^{+0.110}_{-0.134}$
                   & $1.345^{+0.066}_{-0.034}$&
                   & $1.430^{+0.051}_{-0.080}$
                   & $1.581^{+0.124}_{-0.124}$
                   & $1.358^{+0.053}_{-0.045}$&
                   & $1.414^{+0.065}_{-0.062}$
                   & $1.602^{+0.110}_{-0.143}$
                   & $1.357^{+0.053}_{-0.044}$\\

$\beta_0$          & $1.430^{+0.289}_{-0.189}$  %
                   & $2.028^{+0.220}_{-0.191}$   %
                   & $1.480^{+0.256}_{-0.259}$ &  %
                   & $1.503^{+0.228}_{-0.280}$
                   & $2.050^{+0.181}_{-0.198}$
                   & $1.488^{+0.235}_{-0.265}$&
                   & $1.469^{+0.254}_{-0.246}$
                   & $2.002^{+0.225}_{-0.167}$
                   & $1.462^{+0.288}_{-0.236}$&
                   & $1.526^{+0.198}_{-0.293}$
                   & $2.052^{+0.179}_{-0.218}$
                   & $1.475^{+0.259}_{-0.256}$\\

$\beta_1$          & $5.119^{+0.586}_{-0.771}$   %
                   & $3.721^{+0.610}_{-0.550}$   %
                   & $5.168^{+0.771}_{-0.617}$&  %
                   & $4.992^{+0.678}_{-0.625}$
                   & $3.696^{+0.585}_{-0.465}$
                   & $5.199^{+0.686}_{-0.685}$&
                   & $5.028^{+0.646}_{-0.689}$
                   & $3.805^{+0.507}_{-0.629}$
                   & $5.183^{+0.672}_{-0.683}$&
                   & $4.897^{+0.796}_{-0.517}$
                   & $3.707^{+0.601}_{-0.501}$
                   & $5.218^{+0.652}_{-0.676}$\\

$\Omega_{c0}$      & $0.240^{+0.008}_{-0.009}$  %
                   & $0.237^{+0.010}_{-0.009}$  %
                   & $0.236^{+0.011}_{-0.007}$ &  %
                   & $0.237^{+0.011}_{-0.007}$
                   & $0.238^{+0.009}_{-0.009}$
                   & $0.238^{+0.009}_{-0.010}$&
                   & $0.239^{+0.009}_{-0.008}$
                   & $0.235^{+0.012}_{-0.006}$
                   & $0.237^{+0.010}_{-0.009}$&
                   & $0.241^{+0.008}_{-0.010}$
                   & $0.236^{+0.010}_{-0.007}$
                   & $0.236^{+0.010}_{-0.007}$\\

$\Omega_{b0}$      & $0.045^{+0.002}_{-0.002}$  %
                   & $0.045^{+0.002}_{-0.001}$   %
                   & $0.045^{+0.002}_{-0.001}$ &   %
                   & $0.046^{+0.002}_{-0.002}$
                   & $0.045^{+0.002}_{-0.002}$
                   & $0.045^{+0.002}_{-0.002}$&
                   & $0.045^{+0.002}_{-0.001}$
                   & $0.045^{+0.002}_{-0.001}$
                   & $0.045^{+0.002}_{-0.002}$&
                   & $0.045^{+0.002}_{-0.002}$
                   & $0.045^{+0.002}_{-0.002}$
                   & $0.045^{+0.002}_{-0.001}$\\

  $\Omega_{r0}$    & $0.000084$
                   & $0.000083$
                   & $0.000083$          &
                   & $0.000085$%ok
                   & $0.000084$%ok
                   & $0.000086$&%ok
                   & $0.000104$%ok
                   & $0.000087$%ok
                   & $0.000100$&%ok
                   & $0.000102$%ok
                   & $0.000097$%ok
                   & $0.000091$\\ %ok

$\gamma$           & $..........$
                   & $..........$
                   & $..........$                &
                   & $0.0008^{+0.0020}_{-0.0027}$
                   & $-0.0001^{+0.0026}_{-0.0024}$
                   & $0.0001^{+0.0024}_{-0.0025}$&
                   & $0.0024^{+0.0049}_{-0.0063}$
                   & $0.0008^{+0.0052}_{-0.0061}$
                   & $0.0013^{+0.0047}_{-0.0068}$&
                   & $0.0017^{+0.0145}_{-0.0100}$
                   & $0.0039^{+0.0094}_{-0.0145}$
                   & $-0.0002^{+0.0136}_{-0.0119}$\\

$w$                & $-1.051^{+0.049}_{-0.041}$  %
                   & $-1.066^{+0.053}_{-0.043}$  %
                   & $-1.068^{+0.047}_{-0.040}$ &  %
                   & $-1.039^{+0.050}_{-0.056}$
                   & $-1.062^{+0.058}_{-0.048}$
                   & $-1.066^{+0.058}_{-0.051}$&
                   & $-1.038^{+0.058}_{-0.046}$
                   & $-1.062^{+0.065}_{-0.051}$
                   & $-1.050^{+0.048}_{-0.068}$&
                   & $-1.043^{+0.065}_{-0.045}$
                   & $-1.055^{+0.054}_{-0.058}$
                   & $-1.067^{+0.060}_{-0.053}$\\

$h$                & $0.701^{+0.012}_{-0.013}$  %
                   & $0.706^{+0.011}_{-0.014}$  %
                   & $0.707^{+0.010}_{-0.015}$ & %
                   & $0.698^{+0.018}_{-0.018}$
                   & $0.704^{+0.018}_{-0.019}$
                   & $0.733^{+0.017}_{-0.017}$&
                   & $0.701^{+0.011}_{-0.014}$
                   & $0.708^{+0.009}_{-0.016}$
                   & $0.705^{+0.014}_{-0.013}$&
                   & $0.700^{+0.013}_{-0.012}$
                   & $0.705^{+0.012}_{-0.014}$
                   & $0.707^{+0.010}_{-0.014}$\\
\hline
\end{tabular}
\end{table*}

In table \ref{tab:combined_results}, by adopting a linear $\beta$, we give the fitting results of the $w$CDM model and the three IDE models.
From this table we see that, for all the four DE models,
$\beta$ significantly deviates from a constant, consistent with the results of \cite{WWGZ14};
in addition, there is no evidence for the existence of dark sector interaction.
Although the best-fit results of $w$ given by the three LCF are always less than $-1$,
$w=-1$ is still consistent with the current cosmological observations at 2$\sigma$ CL.
Moreover, we check the impacts of different SNLS3 LCF on parameter estimation and find that for all the four DE models:
(1) The ``Combined'' sample always gives the largest $w$; in addition, the values of $w$ given by the ``SALT2'' and the ``SIFTO'' sample are close to each other.
(2) The effects of different LCF on other parameters are negligible.
It is clear that these results are insensitive to a specific DE model.

\begin{figure*}
  \centering
  \resizebox{0.77\columnwidth}{!}{\includegraphics{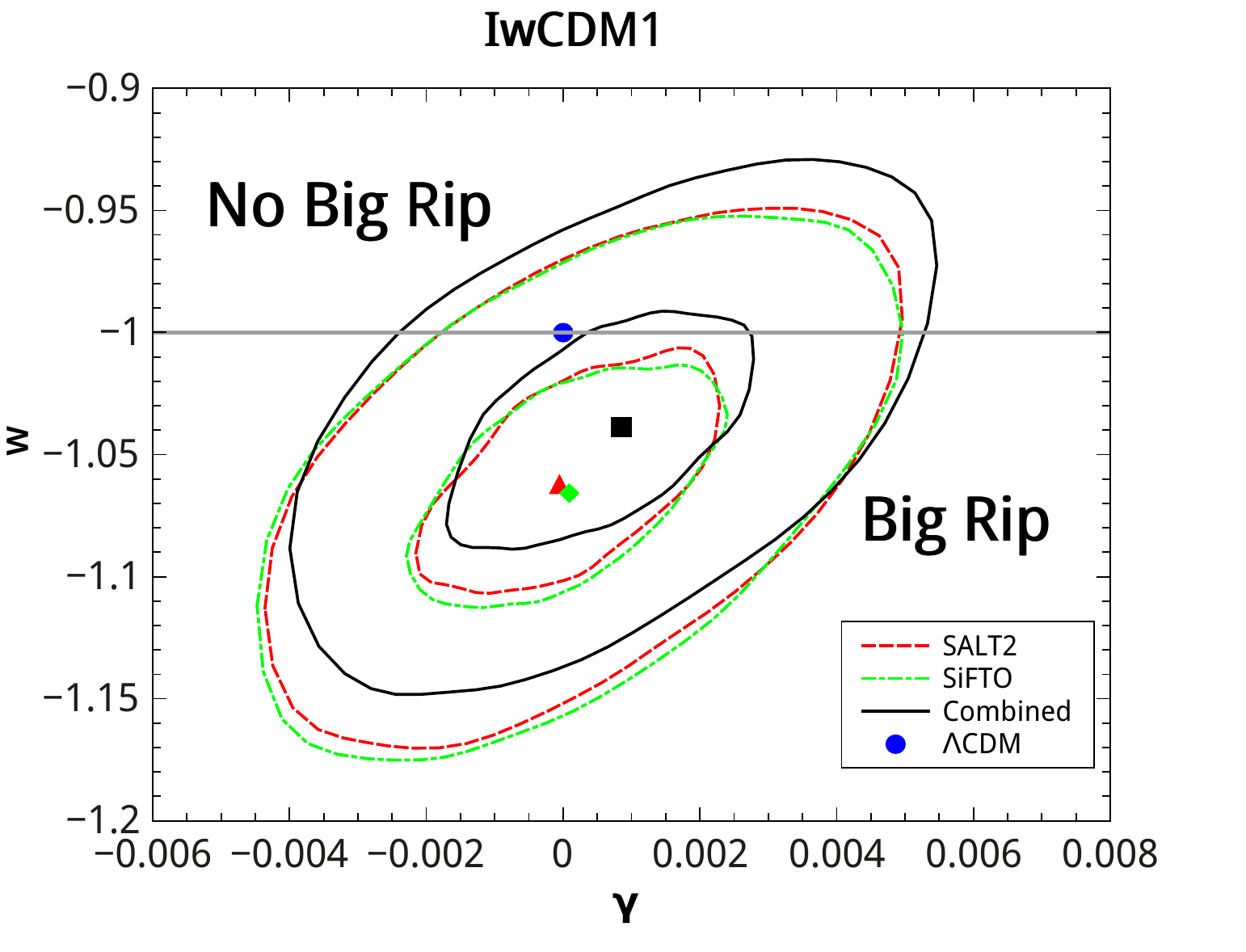}}
  \hspace{0.1\columnwidth}
  \resizebox{0.77\columnwidth}{!}{\includegraphics{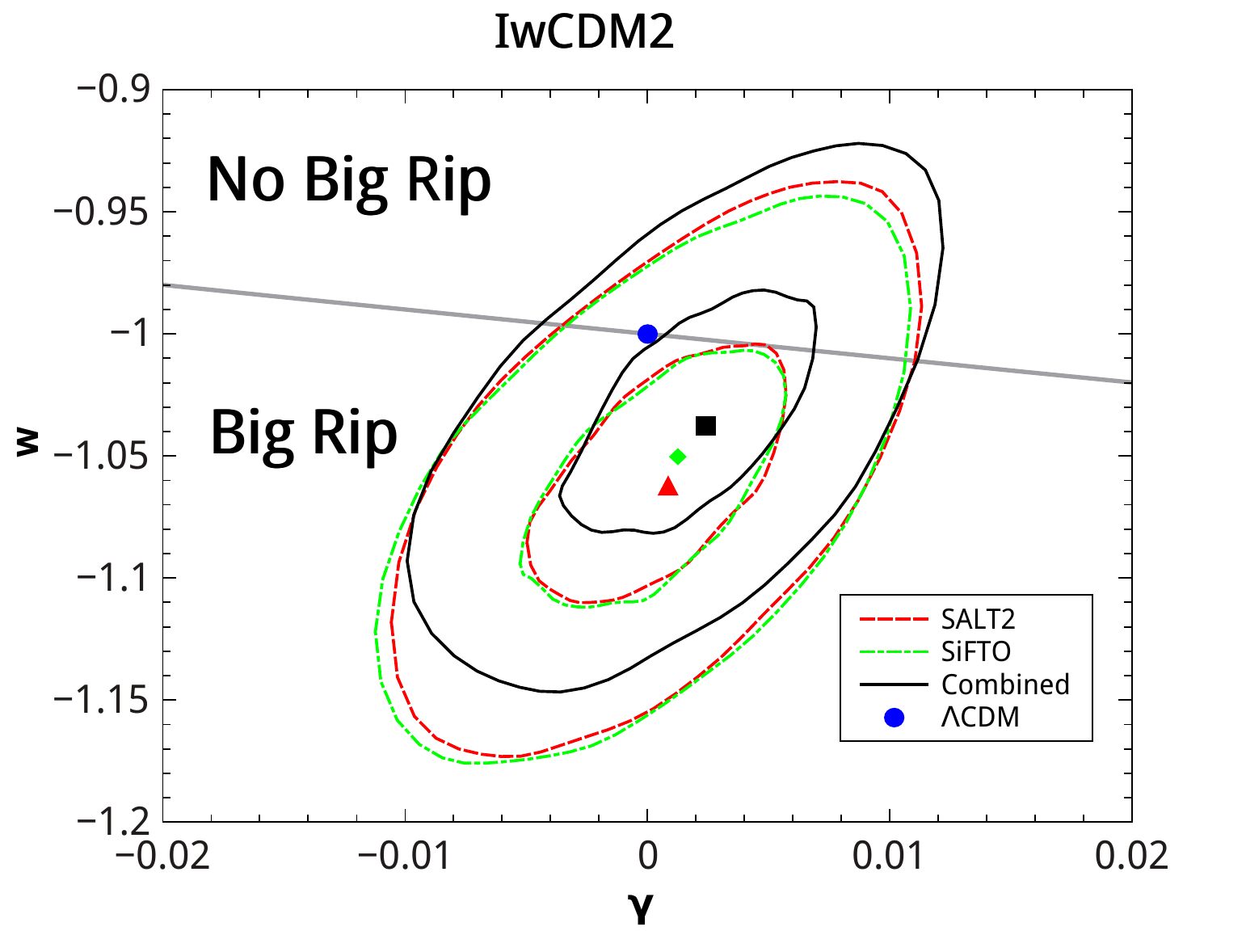}}
   \hspace{0.1\columnwidth}
  \resizebox{0.77\columnwidth}{!}{\includegraphics{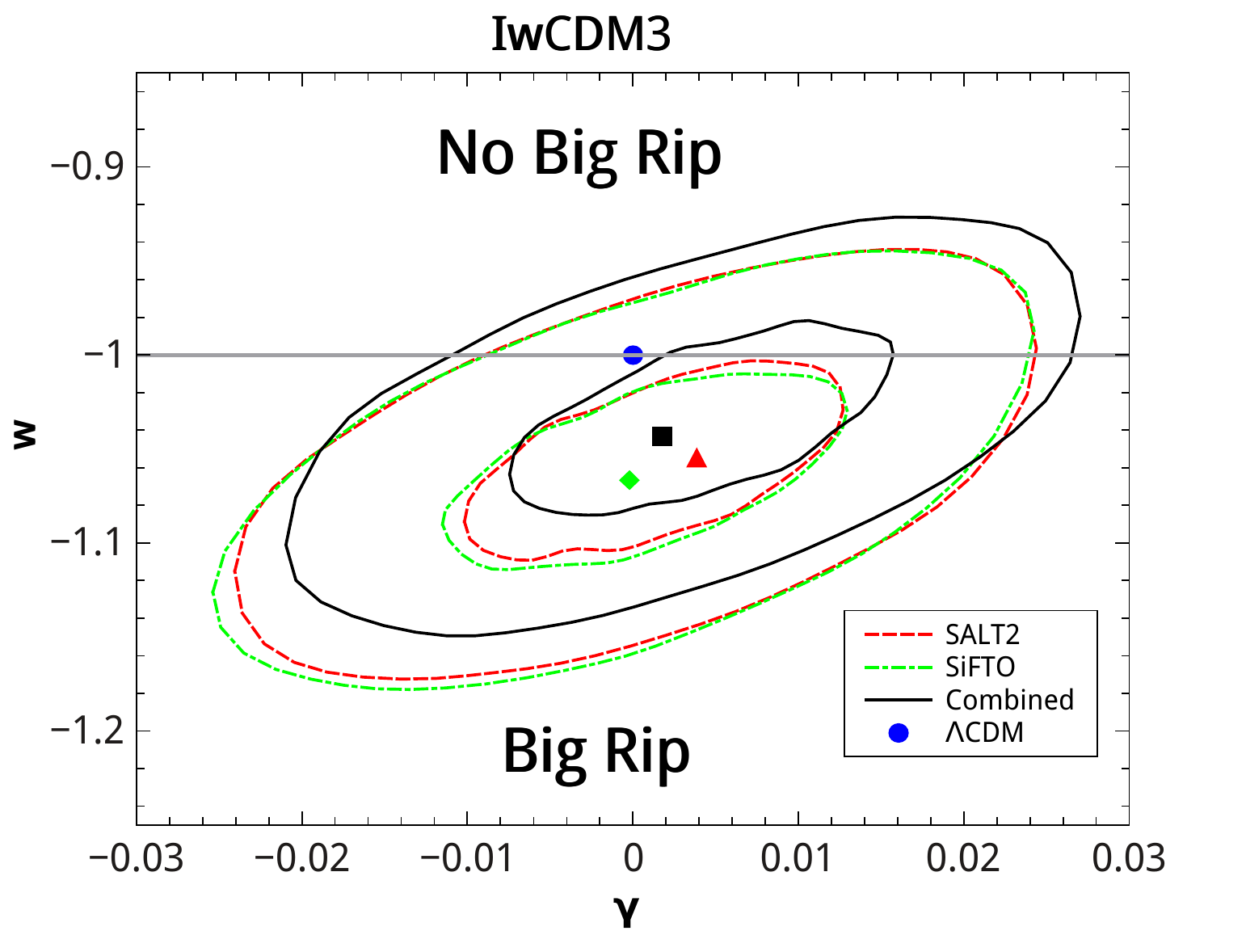}}
\caption{(color online). Probability contours at the 1$\sigma$ and 2$\sigma$ CL in the $\gamma$-$w$ plane,
for the I$w$CDM1 (upper left panel), the I$w$CDM2 (upper right panel) and the I$w$CDM3 (lower panel) model.
``Combined'' (black solid lines), ``SALT2'' (red dashed lines), and ``SiFTO'' (green dash-dotted lines) denote
the results given by the SN(Combined)+CMB+GC, the SN(SALT2)+CMB+GC, and the SN(SiFTO)+CMB+GC data, respectively.
Furthermore, the best-fit values of $\{\gamma, w\}$ of the ``Combined'', the ``SALT2'' and the ``SiFTO'' data
are marked as a black square, a red triangle and a green diamond, respectively.
To make a comparison, the fixed point $\{\gamma, w\} = \{0,-1\}$ for the $\Lambda$CDM model are also marked as a blue round dot.
The gray solid line divides the panel into two regions:
the region above the dividing line denotes a quintessence dominated Universe (without big rip),
and the region below the dividing line represents a phantom dominated Universe (with big rip).}
\label{fig:gamma_w}
\end{figure*}

In Fig. \ref{fig:gamma_w},
we plot the probability contours at the 1$\sigma$ and 2$\sigma$ CL in the $\gamma$-$w$ plane, for the three IDE models.
The best-fit values of $\{\gamma, w\}$ of the ``Combined'', the ``SALT2'' and the ``SiFTO'' data
are marked as a black square, a red triangle and a green diamond, respectively.
For comparison, the fixed point $\{\gamma, w\} = \{0,-1\}$ for the $\Lambda$CDM model is also marked as a blue round dot.
A most obvious feature of this figure is that, for all the IDE models,
the fixed point $ \{0,-1\}$ of the $\Lambda$CDM model is outside the 1$\sigma$ contours given by the three data sets;
however, the $\Lambda$CDM model is still consistent with the observational data at 2$\sigma$ CL.
Moreover, according to the evolution behaviors of $\rho_{de}$ (see Eqs. \ref{rho:de1}, \ref{rho:de2} and \ref{rho:de3}) at $z \rightarrow -1$,
we divide these $\gamma$-$w$ planes into two regions:
the region above the dividing line denotes a quintessence dominated Universe (without big rip),
and the region below the dividing line represents a phantom dominated Universe (with big rip).
We can see that, although all the best-fit points given by the three data sets correspond to a phantom,
both phantom, quintessence and cosmological constant
are consistent with the current cosmological observations at 2$\sigma$ CL.
This means that the current observational data are still too limited to indicate the nature of DE.

\subsection{Hubble Parameter, Deceleration Parameter and Statefinder Hierarchy}

\begin{figure*}
  \centering
  \resizebox{0.77\columnwidth}{!}{\includegraphics{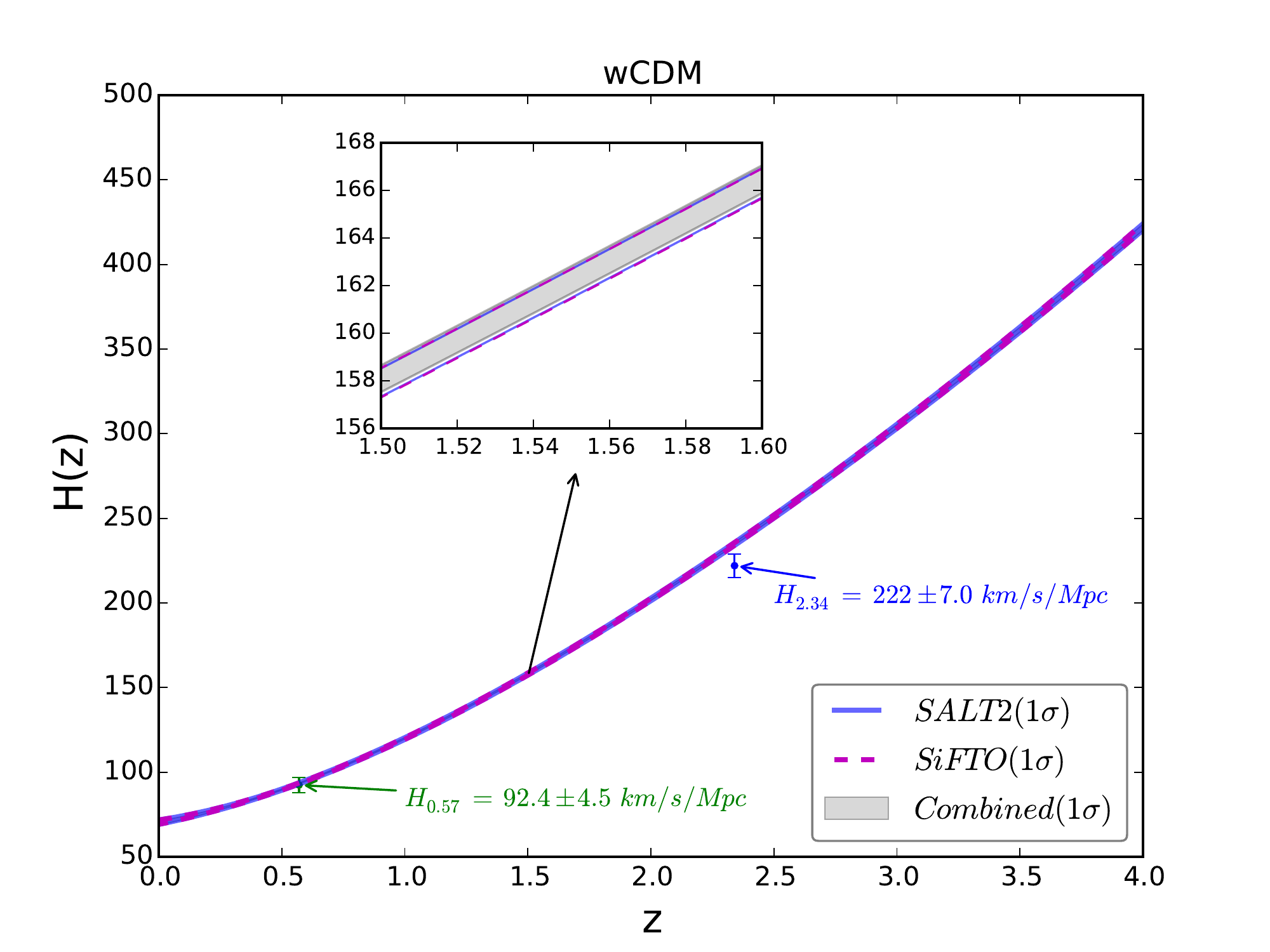}}
  \hspace{0.1\columnwidth}
  \resizebox{0.77\columnwidth}{!}{\includegraphics{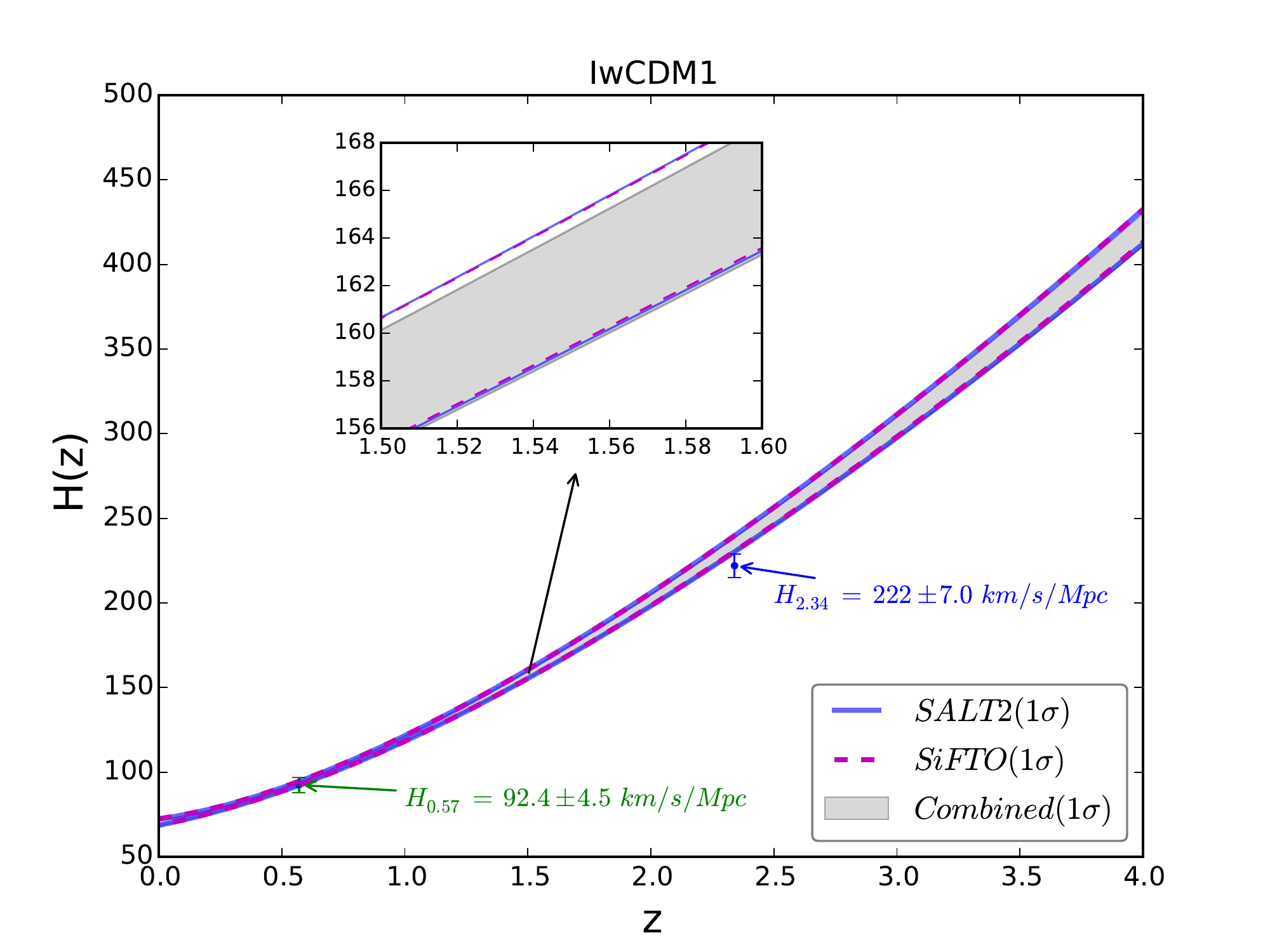}}
   \hspace{0.1\columnwidth}
  \resizebox{0.77\columnwidth}{!}{\includegraphics{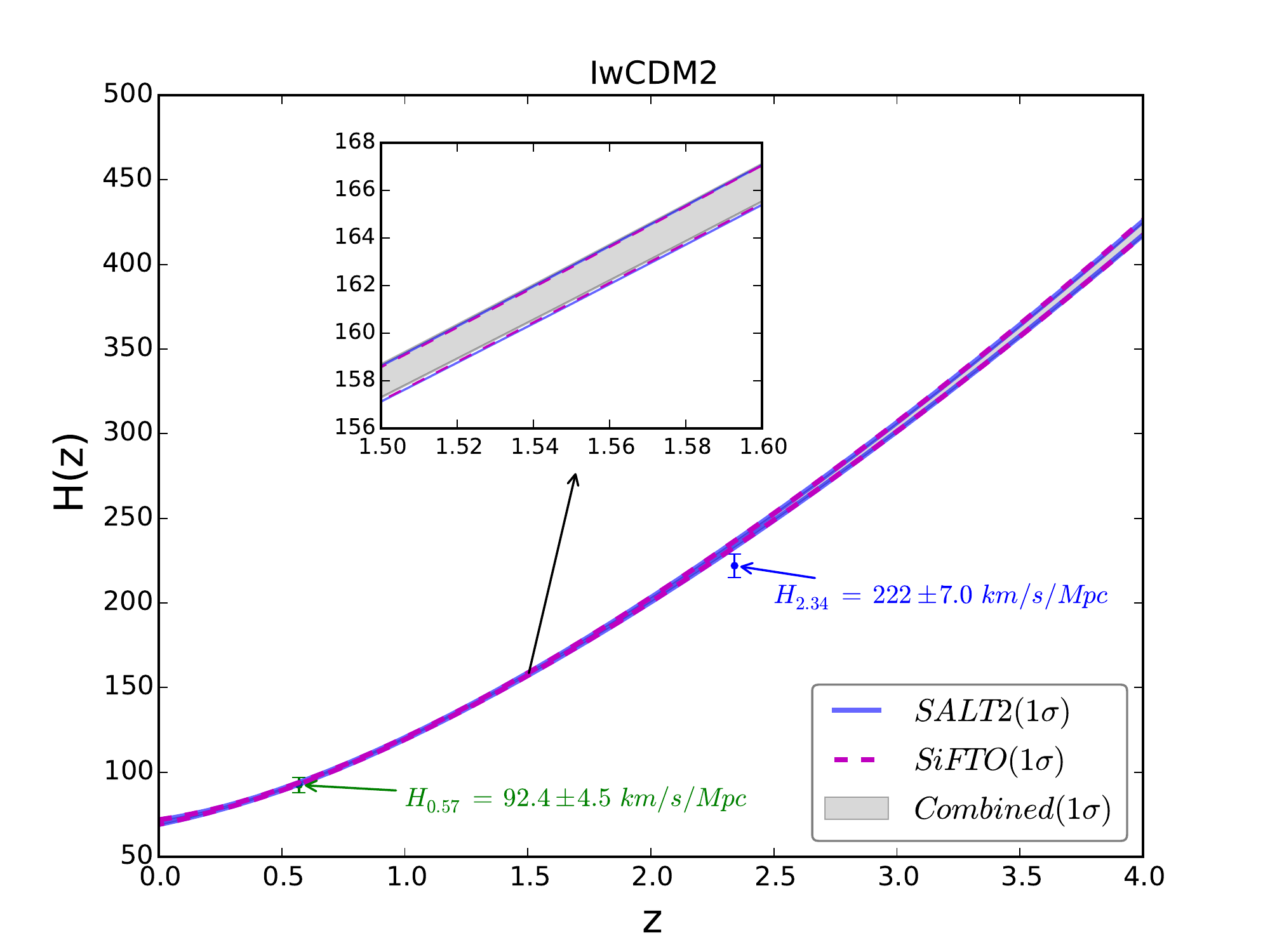}}
   \hspace{0.1\columnwidth}
  \resizebox{0.77\columnwidth}{!}{\includegraphics{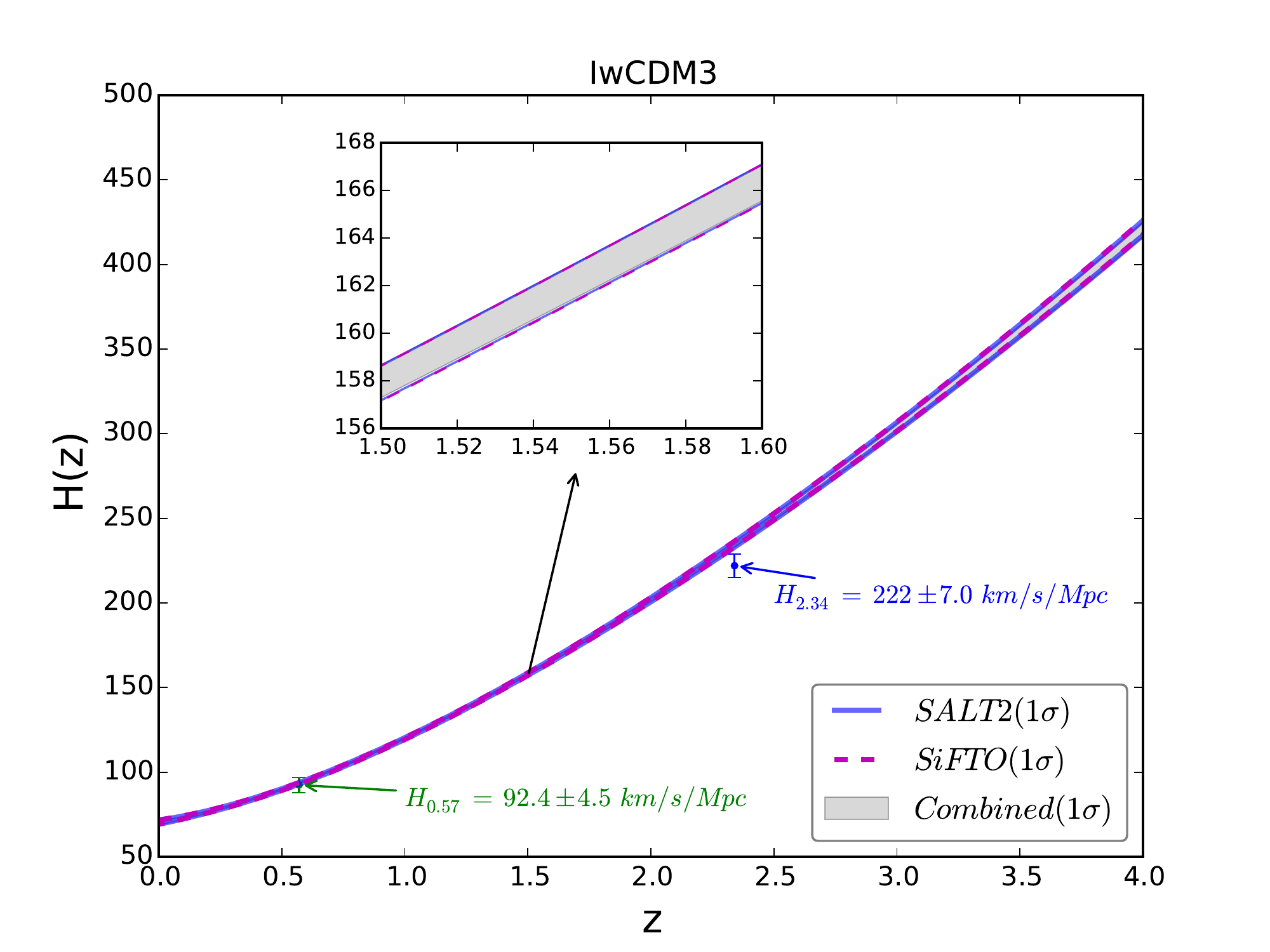}}
\caption{(color online). The 1$\sigma$ confidence regions of Hubble parameter $H(z)$ at redshift region $[0,4]$,
for the $w$CDM (upper left panel), the I$w$CDM1 (upper right panel), the I$w$CDM2 (lower left panel) and the I$w$CDM3 (lower right panel) model, where the data points of $H_{0.57}$ and $H_{2.34}$ are also marked by diamonds with error bars for comparison.
``Combined'' (gray filled regions), ``SALT2'' (blue solid lines), and ``SiFTO'' (purple dashed lines) denote
the results given by the SN(Combined)+CMB+GC, the SN(SALT2)+CMB+GC, and the SN(SiFTO)+CMB+GC data, respectively.
}
\label{fig:hz}
\end{figure*}

The 1$\sigma$ confidence regions of Hubble parameter $H(z)$ at redshift region $[0,4]$ for the $w$CDM model and the three IDE models are plotted in Fig. \ref{fig:hz},
where the two $H(z)$ data points, $H_{0.57}$ and $H_{2.34}$, are also marked by diamonds with error bars for comparison.
We find that the data point $H_{0.57}$ can be easily accommodated in the $w$CDM model and the three IDE models,
but the data point $H_{2.34}$ significantly deviates from the 1$\sigma$ regions of all the four DE models.
In other words, the measurement of $H_{2.34}$ is in tension with other cosmological observations
and this result is consistent with the conclusion of \cite{Sahni14, HLZ2014}.
In addition, the 1$\sigma$ confidence regions of $H(z)$ given by different LCF are almost overlap;
this means that using $H(z)$ diagram is almost impossible to distinguish the differences among different SNLS3 LCF.

\begin{figure*}
  \centering
  \resizebox{0.77\columnwidth}{!}{\includegraphics{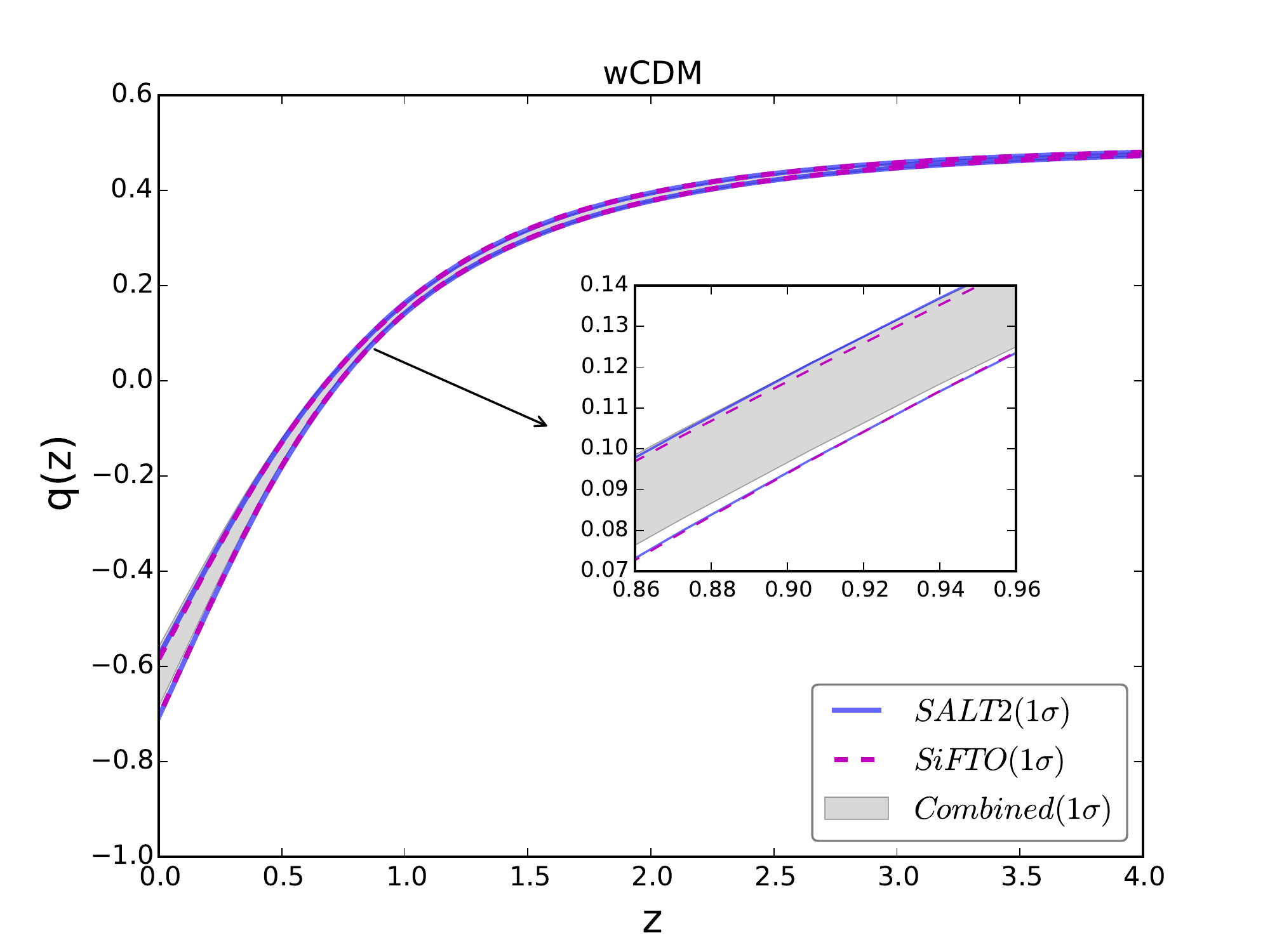}}
  \hspace{0.1\columnwidth}
  \resizebox{0.77\columnwidth}{!}{\includegraphics{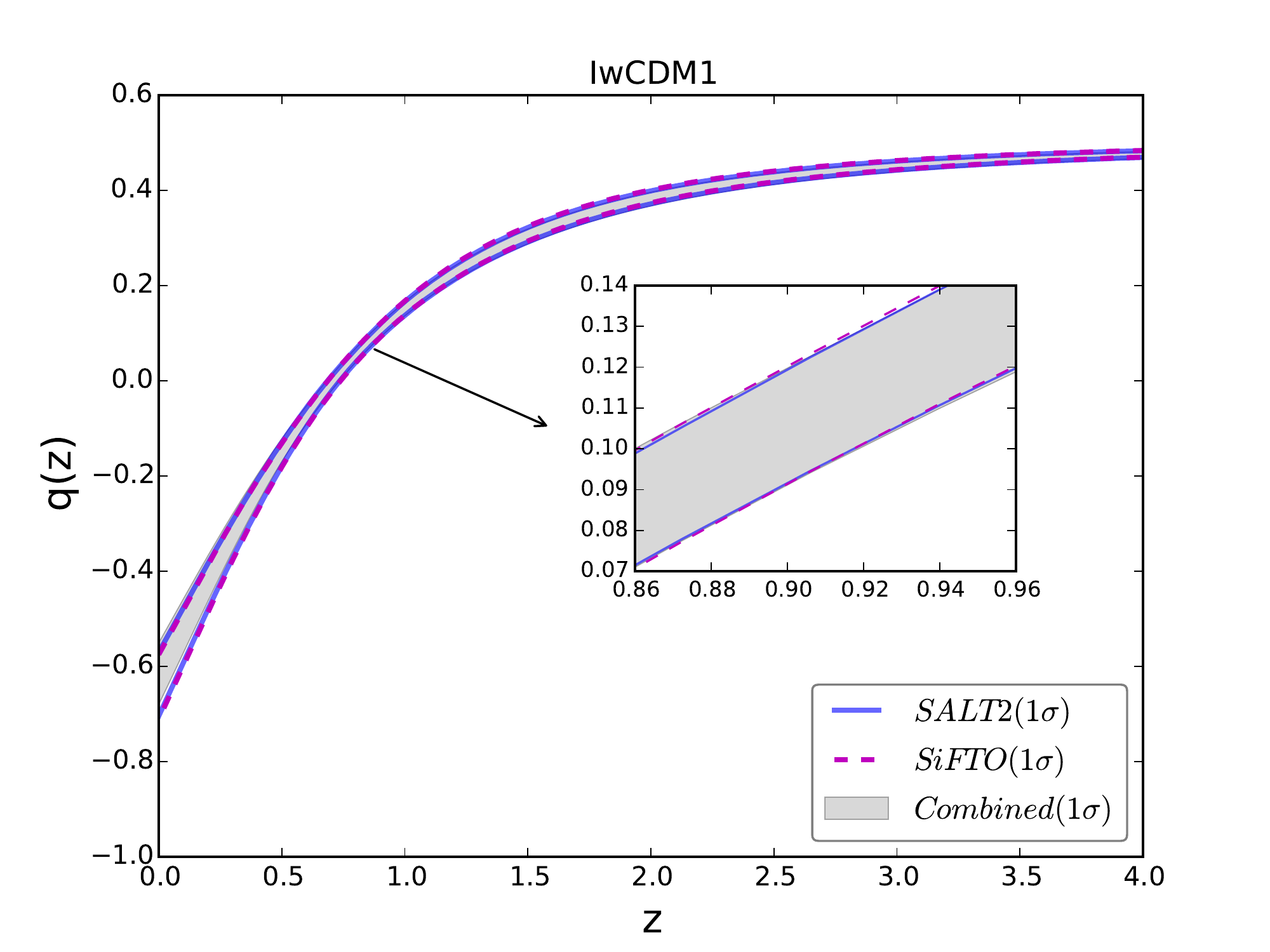}}
   \hspace{0.1\columnwidth}
  \resizebox{0.77\columnwidth}{!}{\includegraphics{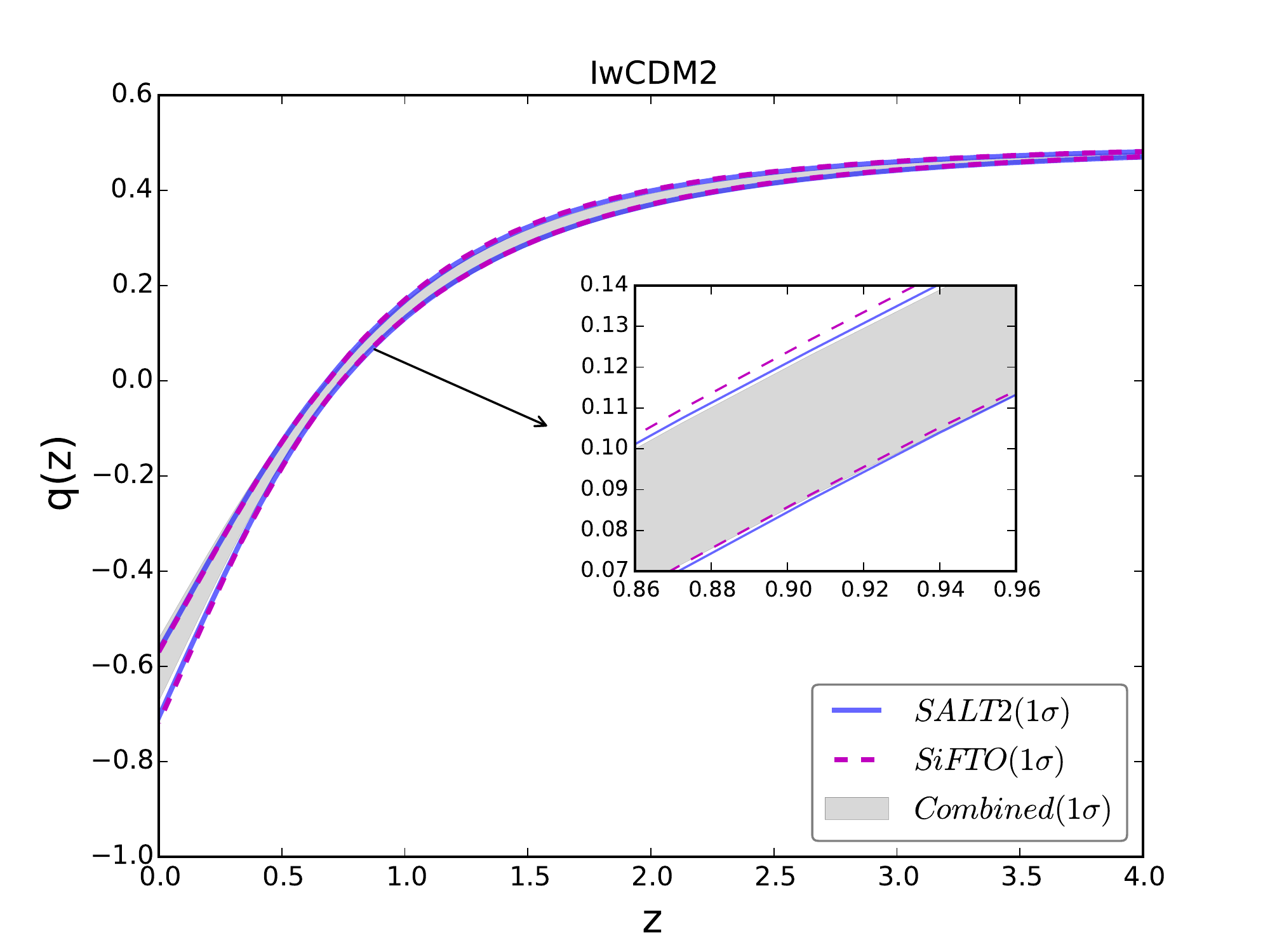}}
   \hspace{0.1\columnwidth}
  \resizebox{0.77\columnwidth}{!}{\includegraphics{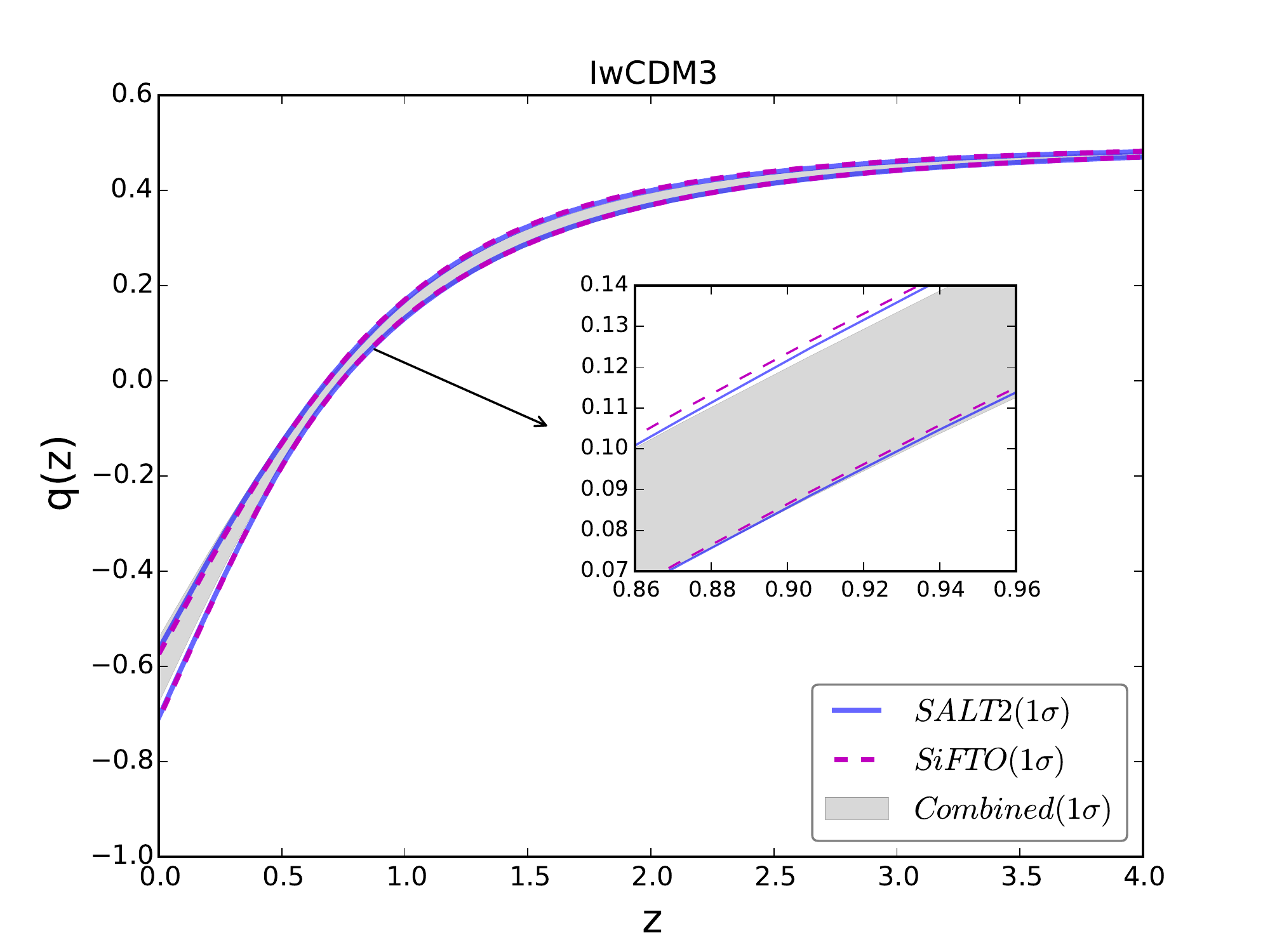}}
\caption{(color online). The 1$\sigma$ confidence regions of deceleration parameter $q(z)$ at redshift region $[0,4]$
for the $w$CDM (upper left panel), the I$w$CDM1 (upper right panel), the I$w$CDM2 (lower left panel) and the I$w$CDM3 (lower right panel) model.
``Combined'' (gray filled regions), ``SALT2'' (blue solid lines), and ``SiFTO'' (purple dashed lines) denote
the results given by the SN(Combined)+CMB+GC, the SN(SALT2)+CMB+GC, and the SN(SiFTO)+CMB+GC data, respectively.
}
\label{fig:qz}
\end{figure*}

In Fig. \ref{fig:qz},
we plot the 1$\sigma$ confidence regions of deceleration parameter $q(z)$ at redshift region $[0,4]$, for the $w$CDM model and the three IDE models.
Again, we see that the 1$\sigma$ confidence regions of $q(z)$ given by different LCF are almost overlap.
This implies that using $q(z)$ diagram also has great difficulty to distinguish the differences among different SNLS3 LCF.

\begin{figure*}
  \centering
  \resizebox{0.77\columnwidth}{!}{\includegraphics{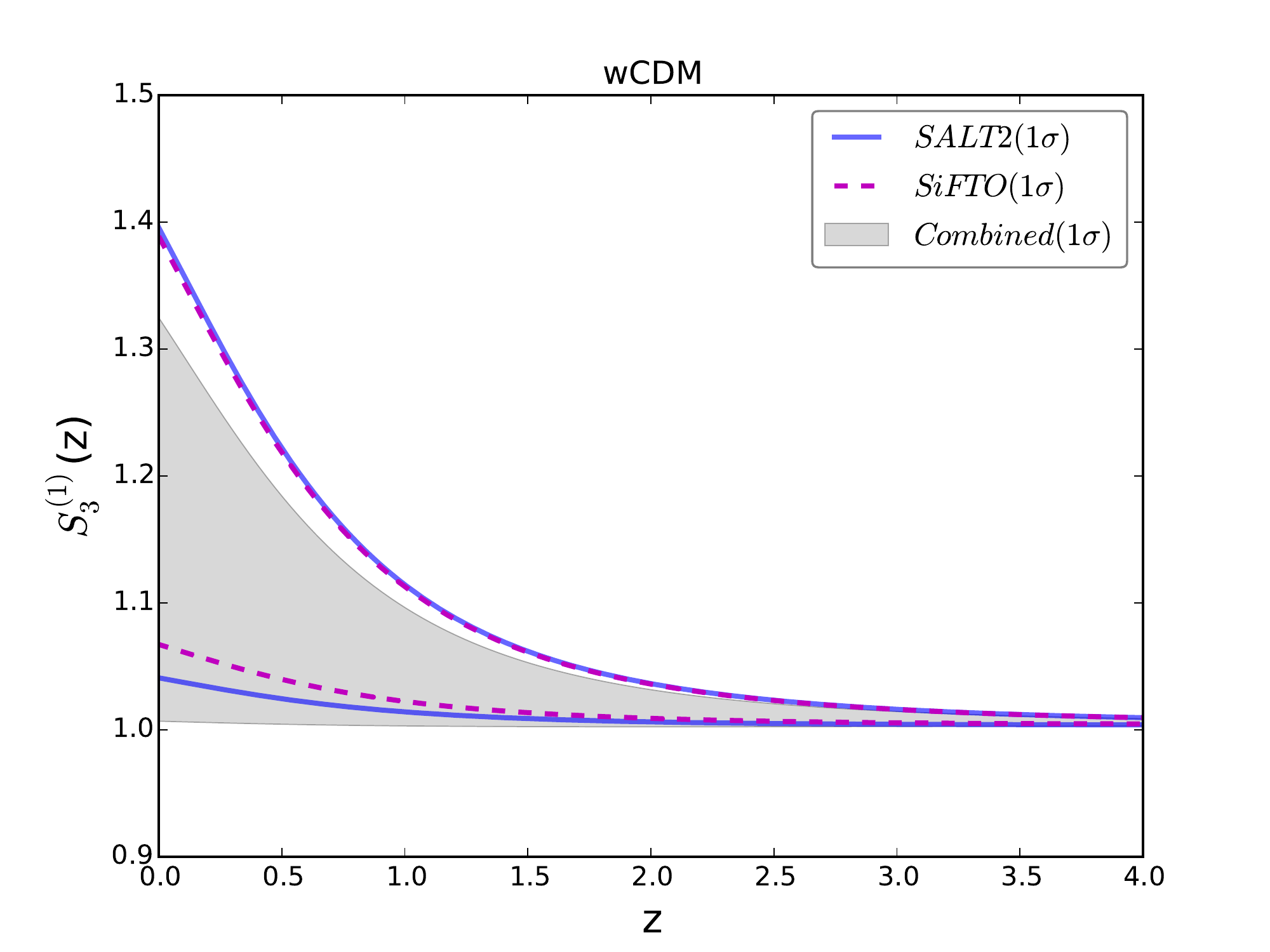}}
  \hspace{0.1\columnwidth}
  \resizebox{0.77\columnwidth}{!}{\includegraphics{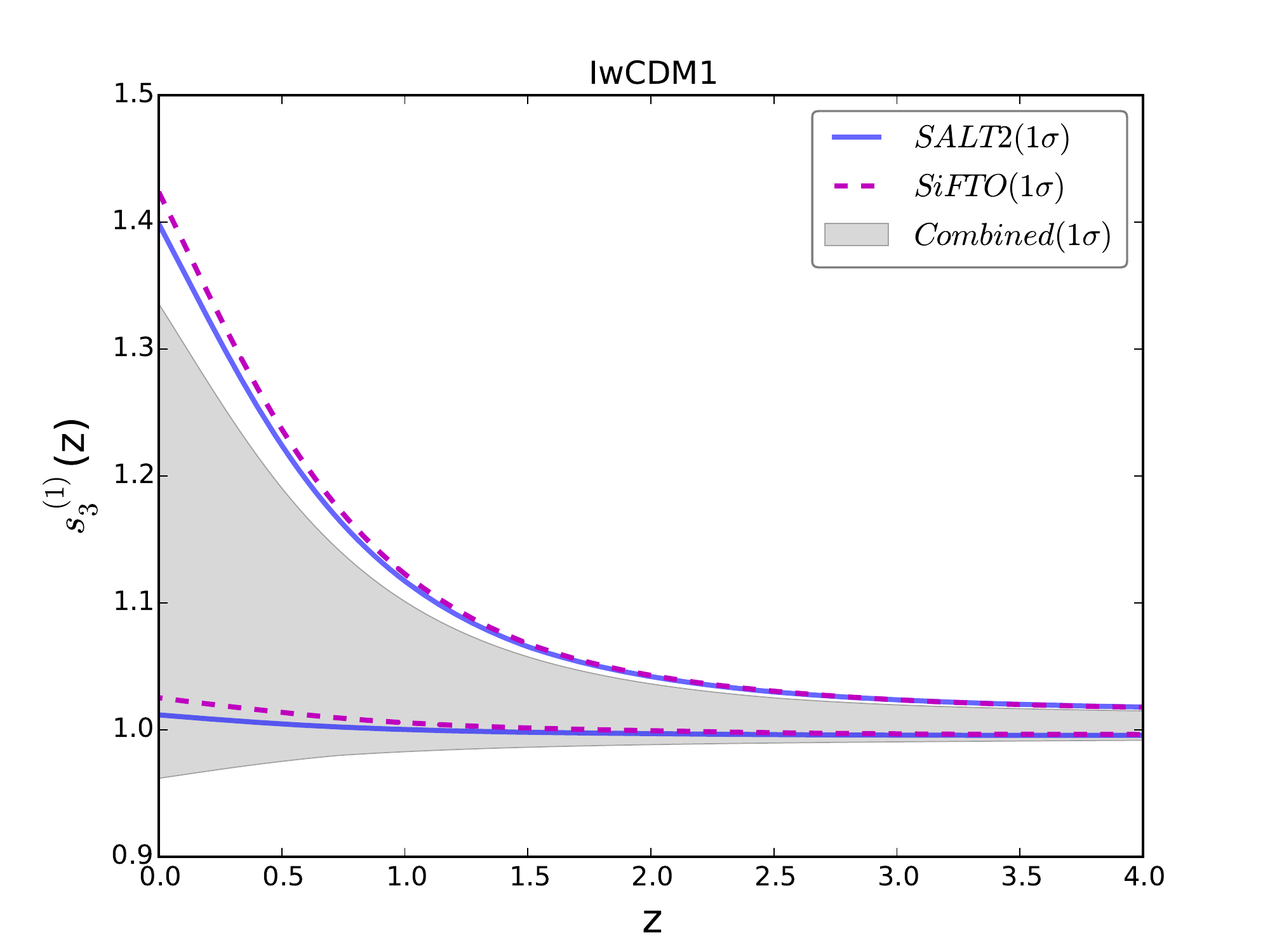}}
   \hspace{0.1\columnwidth}
  \resizebox{0.77\columnwidth}{!}{\includegraphics{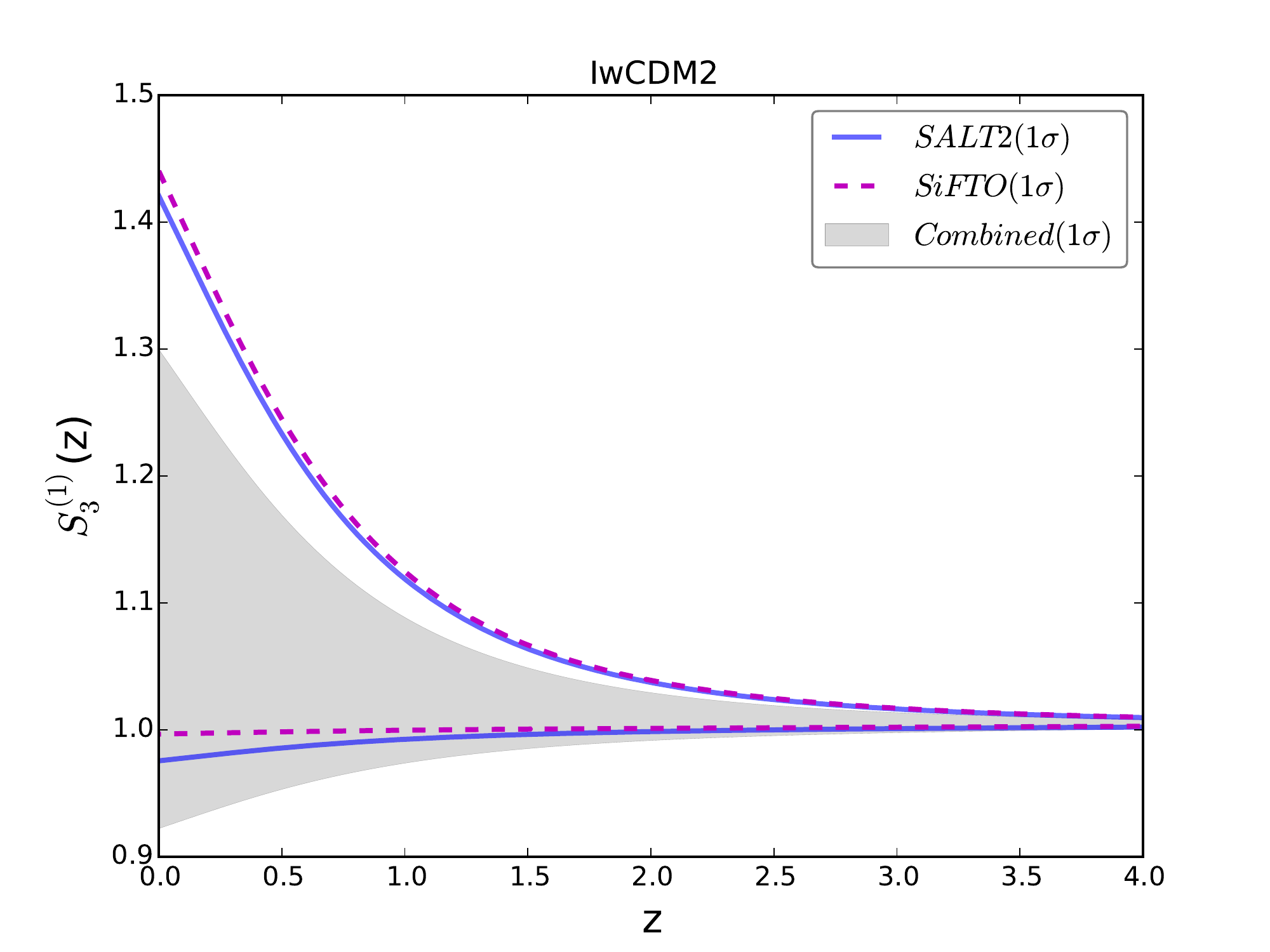}}
   \hspace{0.1\columnwidth}
  \resizebox{0.77\columnwidth}{!}{\includegraphics{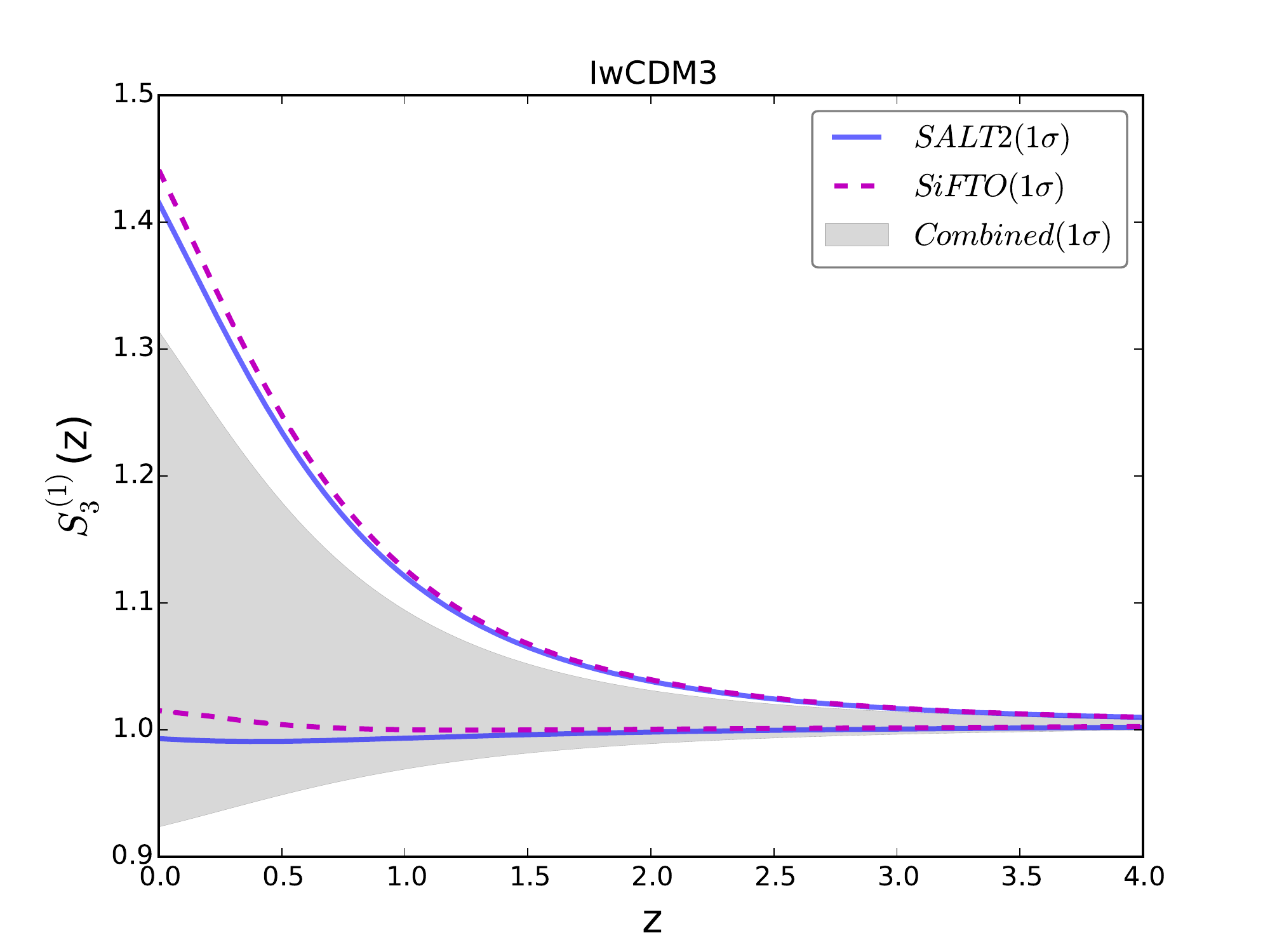}}
\caption{(color online). The 1$\sigma$ confidence regions of $S^{(1)}_3(z)$ at redshift region $[0,4]$
for the $w$CDM (upper left panel), the I$w$CDM1 (upper right panel), the I$w$CDM2 (lower left panel) and the I$w$CDM3 (lower right panel) model.
``Combined'' (gray filled regions), ``SALT2'' (blue solid lines), and ``SiFTO'' (purple dashed lines) denote
the results given by the SN(Combined)+CMB+GC, the SN(SALT2)+CMB+GC, and the SN(SiFTO)+CMB+GC data, respectively.
}
\label{fig:statefinder3}
\end{figure*}

In Fig. \ref{fig:statefinder3},
we plot the 1$\sigma$ confidence regions of $S^{(1)}_3(z)$ at redshift region $[0,4]$, for the $w$CDM model and the three IDE models.
From this figure we see that, the 1$\sigma$ confidence regions of $S^{(1)}_3(z)$
given by the three SNLS3 LCF are almost overlap at high redshift.
Although there is an separating trend for $S^{(1)}_3(z)$ when $z \rightarrow 0$,
most parts of these three 1$\sigma$ regions are overlap even at current epoch.
This means that, although $S^{(1)}_3(z)$ is a better tool than $H(z)$ and $q(z)$,
it still have difficulty to distinguish the effects of different SNLS3 LCF.
Moreover, our conclusion holds true for all the four DE models,
showing that this conclusion is insensitive to a specific interaction form.

\begin{figure*}
  \centering
  \resizebox{0.77\columnwidth}{!}{\includegraphics{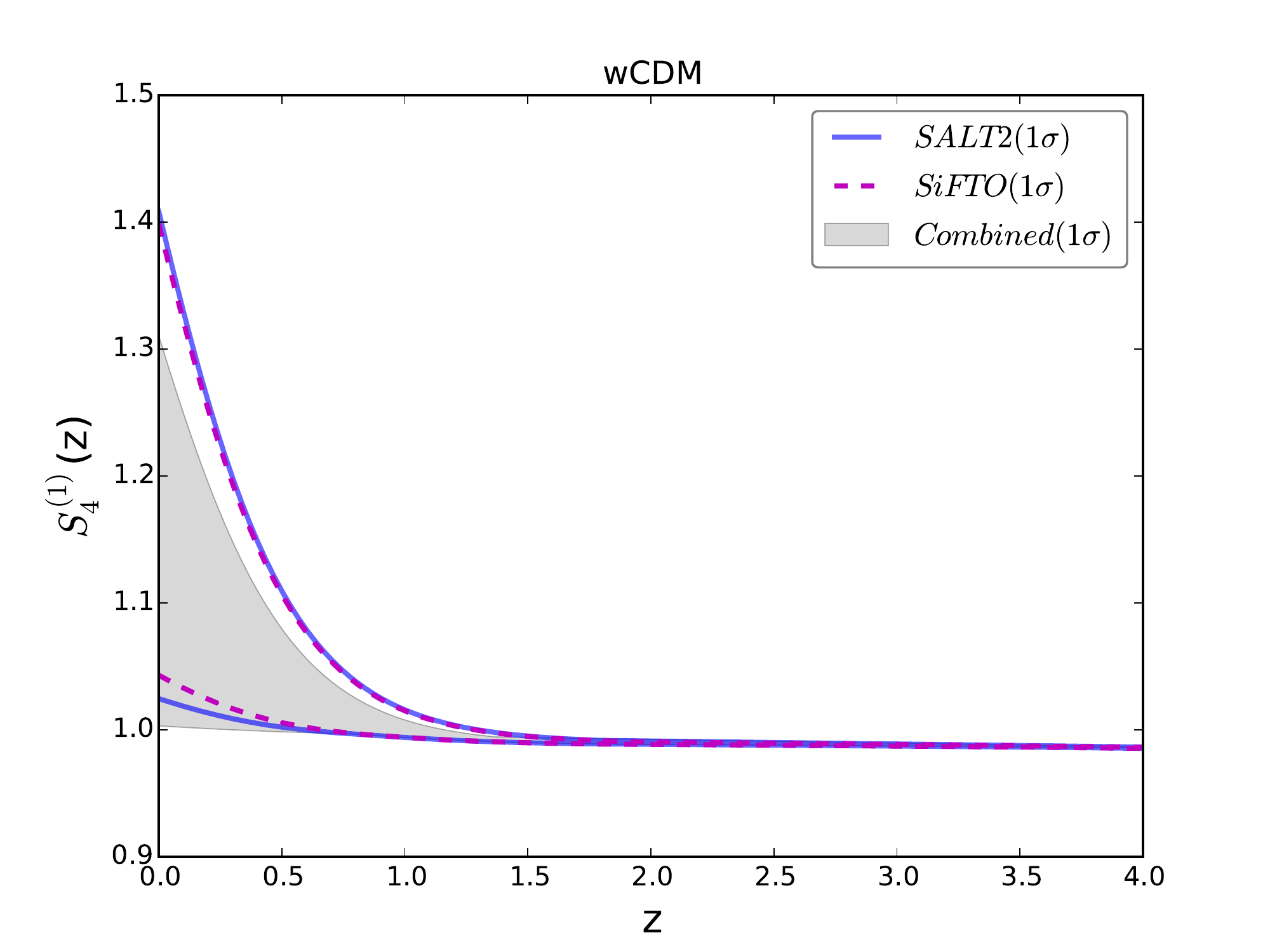}}
  \hspace{0.1\columnwidth}
  \resizebox{0.77\columnwidth}{!}{\includegraphics{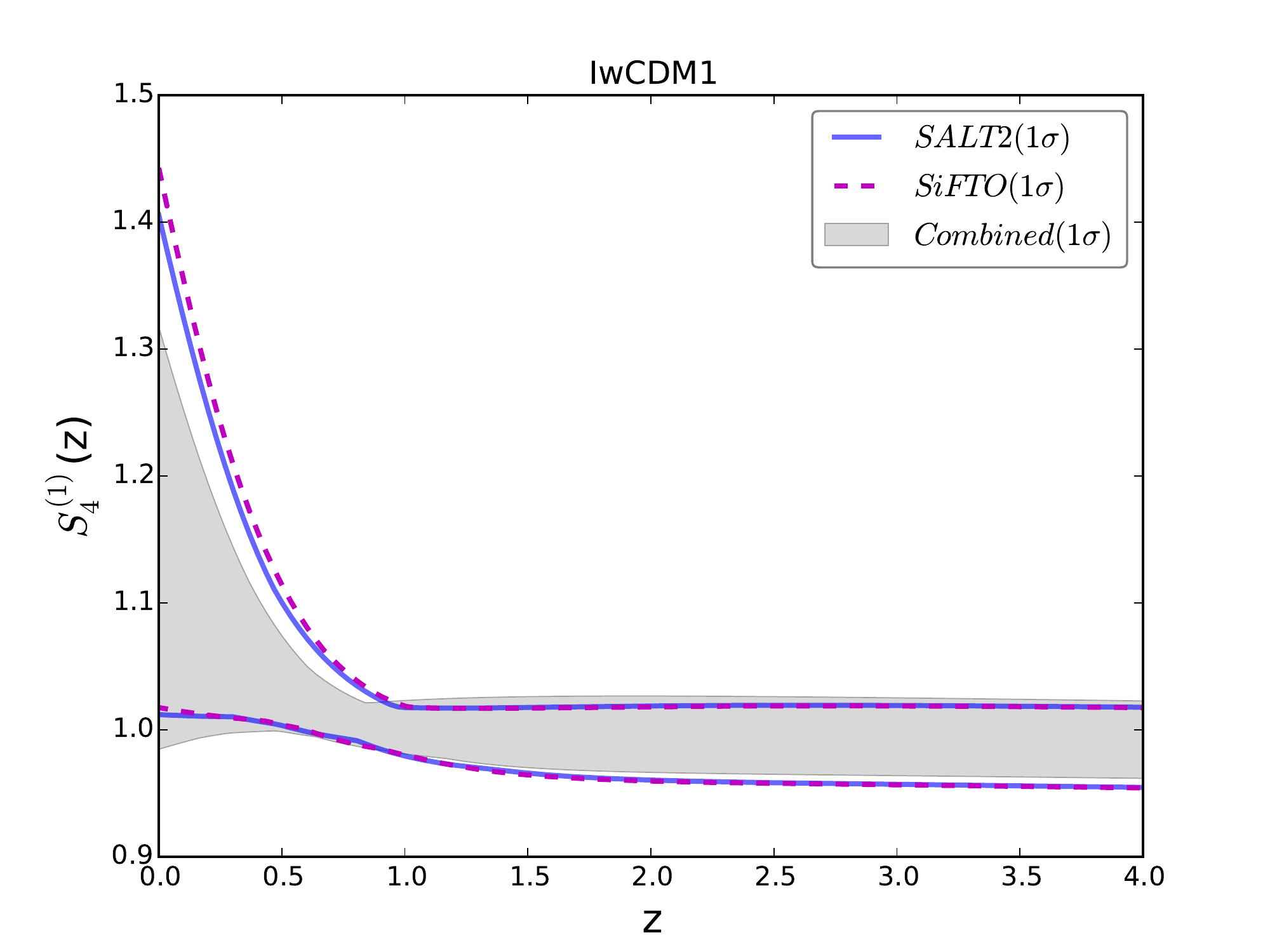}}
   \hspace{0.1\columnwidth}
  \resizebox{0.77\columnwidth}{!}{\includegraphics{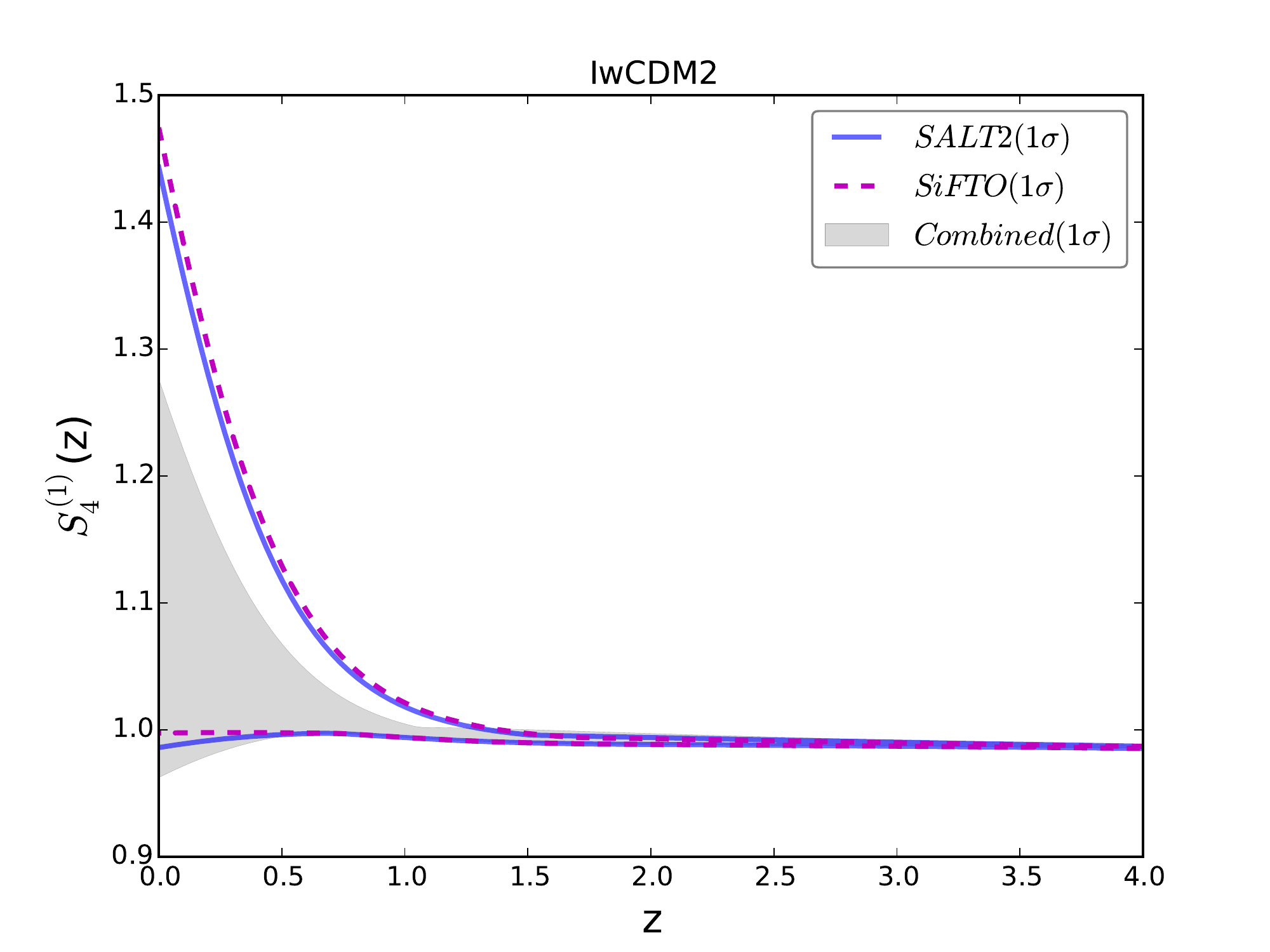}}
   \hspace{0.1\columnwidth}
  \resizebox{0.77\columnwidth}{!}{\includegraphics{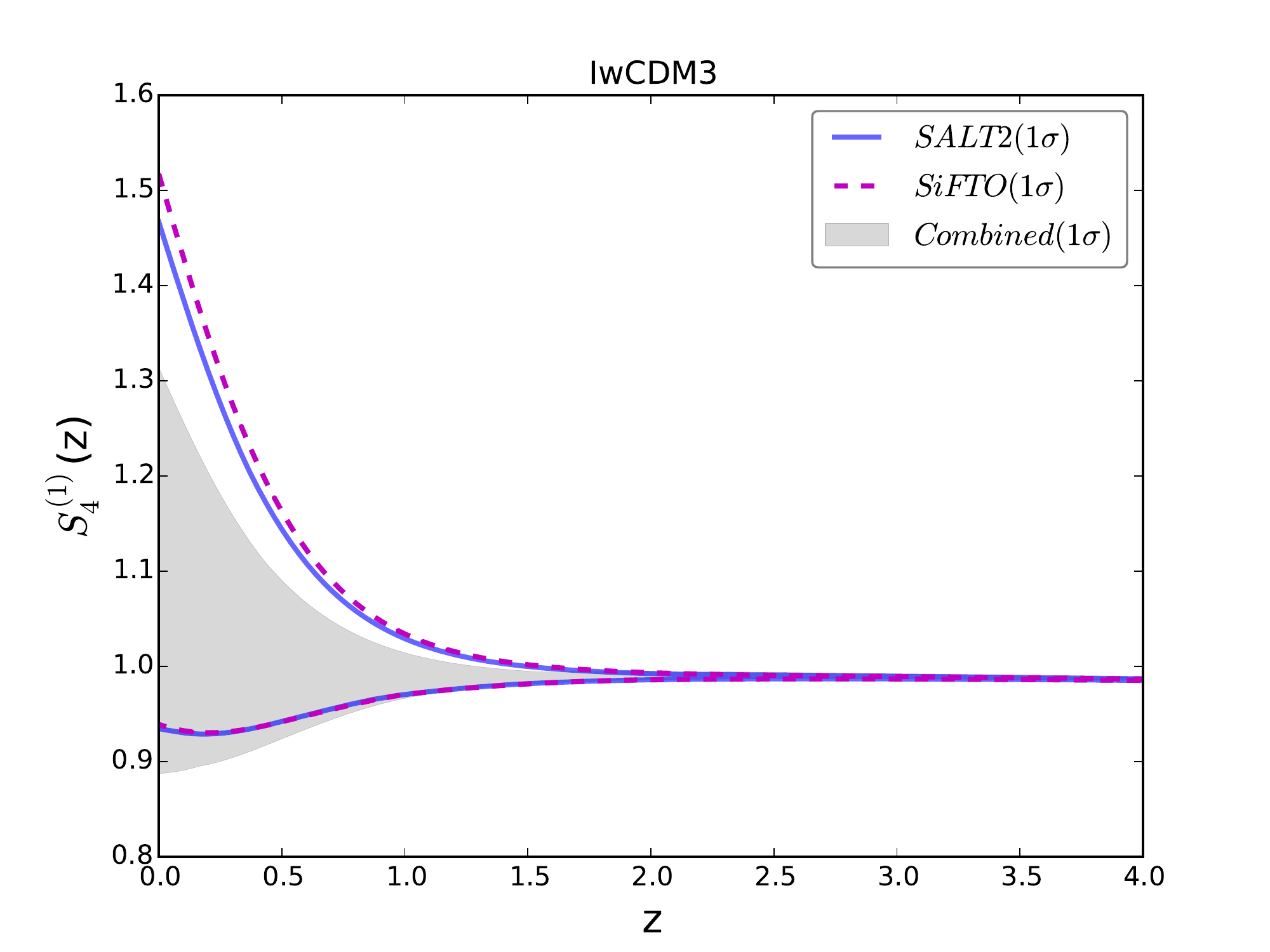}}
\caption{(color online). The 1$\sigma$ confidence regions of $S^{(1)}_4(z)$ at redshift region $[0,4]$
for the $w$CDM (upper left panel), the I$w$CDM1 (upper right panel), the I$w$CDM2 (lower left panel) and the I$w$CDM3 (lower right panel) model.
``Combined'' (gray filled regions), ``SALT2'' (blue solid lines), and ``SiFTO'' (purple dashed lines) denote
the results given by the SN(Combined)+CMB+GC, the SN(SALT2)+CMB+GC, and the SN(SiFTO)+CMB+GC data, respectively.
}
\label{fig:statefinder4}
\end{figure*}

In Fig. \ref{fig:statefinder4},
we plot the 1$\sigma$ confidence regions of $S^{(1)}_4(z)$ at redshift region $[0,4]$, for the $w$CDM model and the three IDE models.
From this figure we see that most of the 1$\sigma$ confidence regions of $S^{(1)}_4(z)$ given by the three LCF are overlap.
This means that the effects of different LCF can not be distinguished by using the statefinder $S^{(1)}_4(z)$ either.
Again, we see that this conclusion is insensitive to a specific interaction form.

In conclusion, we find that the differences given by ``SALT2'', ``SiFTO'', and ``Combined'' LCF are rather small and can not be distinguished by using $H(z)$, $q(z)$, $S^{(1)}_3(z)$ and $S^{(1)}_4(z)$.
This result is quite different from the case of ``MLCS2k2'' \cite{Jha07} and ``SALT2'' \cite{Guy07},
where using ``MLCS2k2'' and ``SALT2'' LCF
will give completely different cosmological constraints for various models \cite{Bengochea11,Bengochea14}.

\subsection{Cosmic Age and Fate of The Universe}

\begin{figure*}
  \centering
  \resizebox{0.77\columnwidth}{!}{\includegraphics{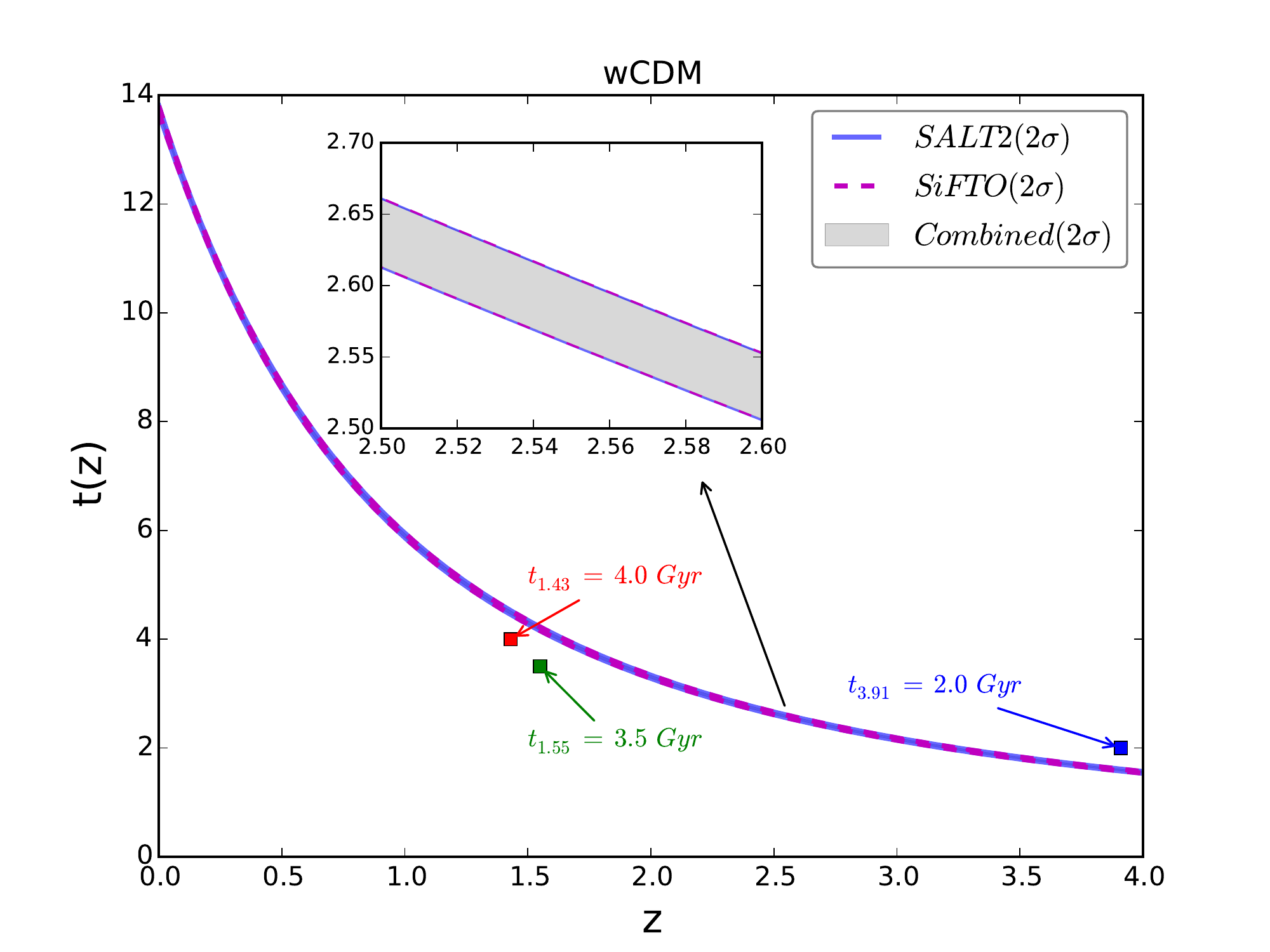}}
  \hspace{0.1\columnwidth}
  \resizebox{0.77\columnwidth}{!}{\includegraphics{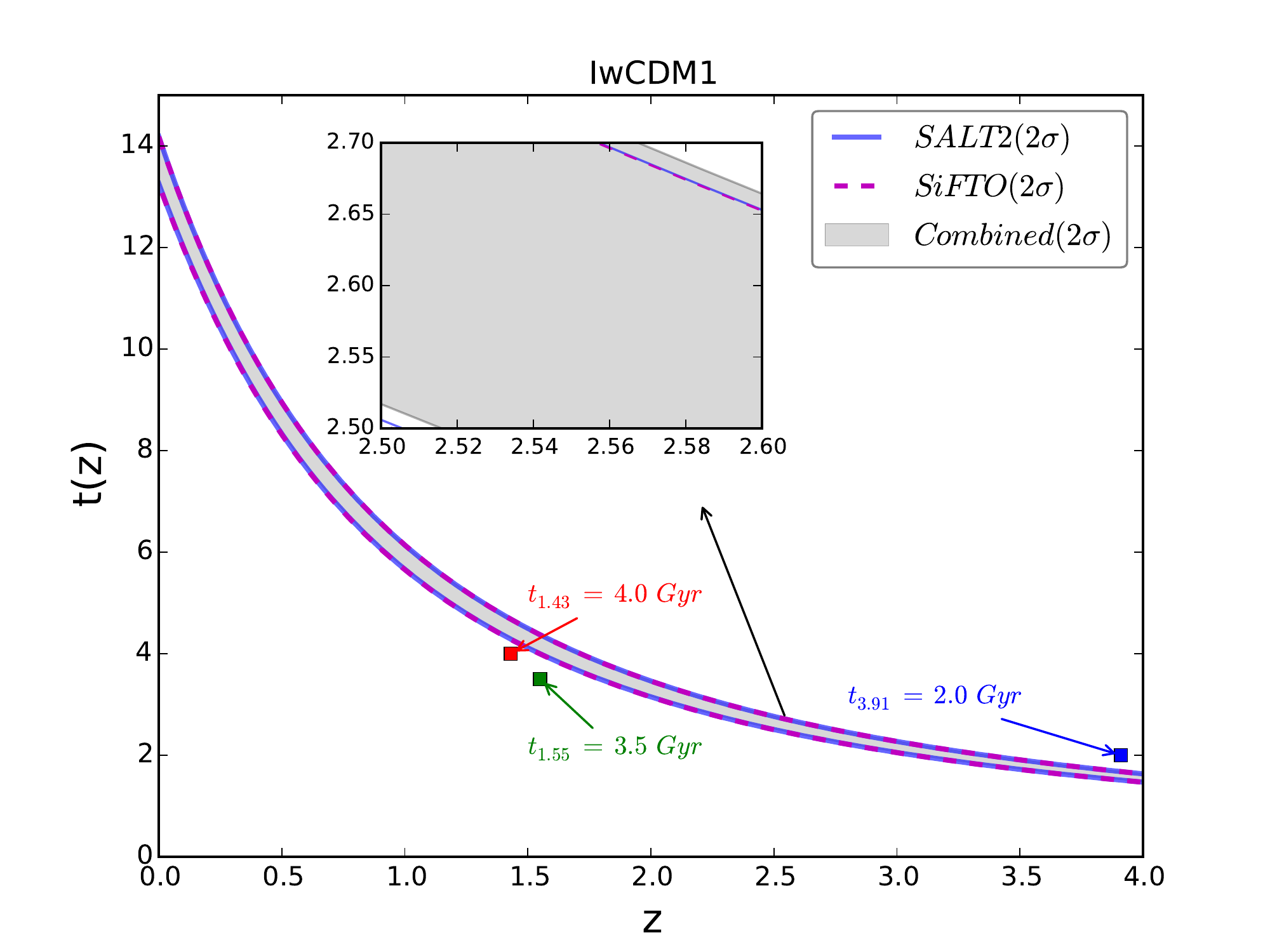}}
   \hspace{0.1\columnwidth}
  \resizebox{0.77\columnwidth}{!}{\includegraphics{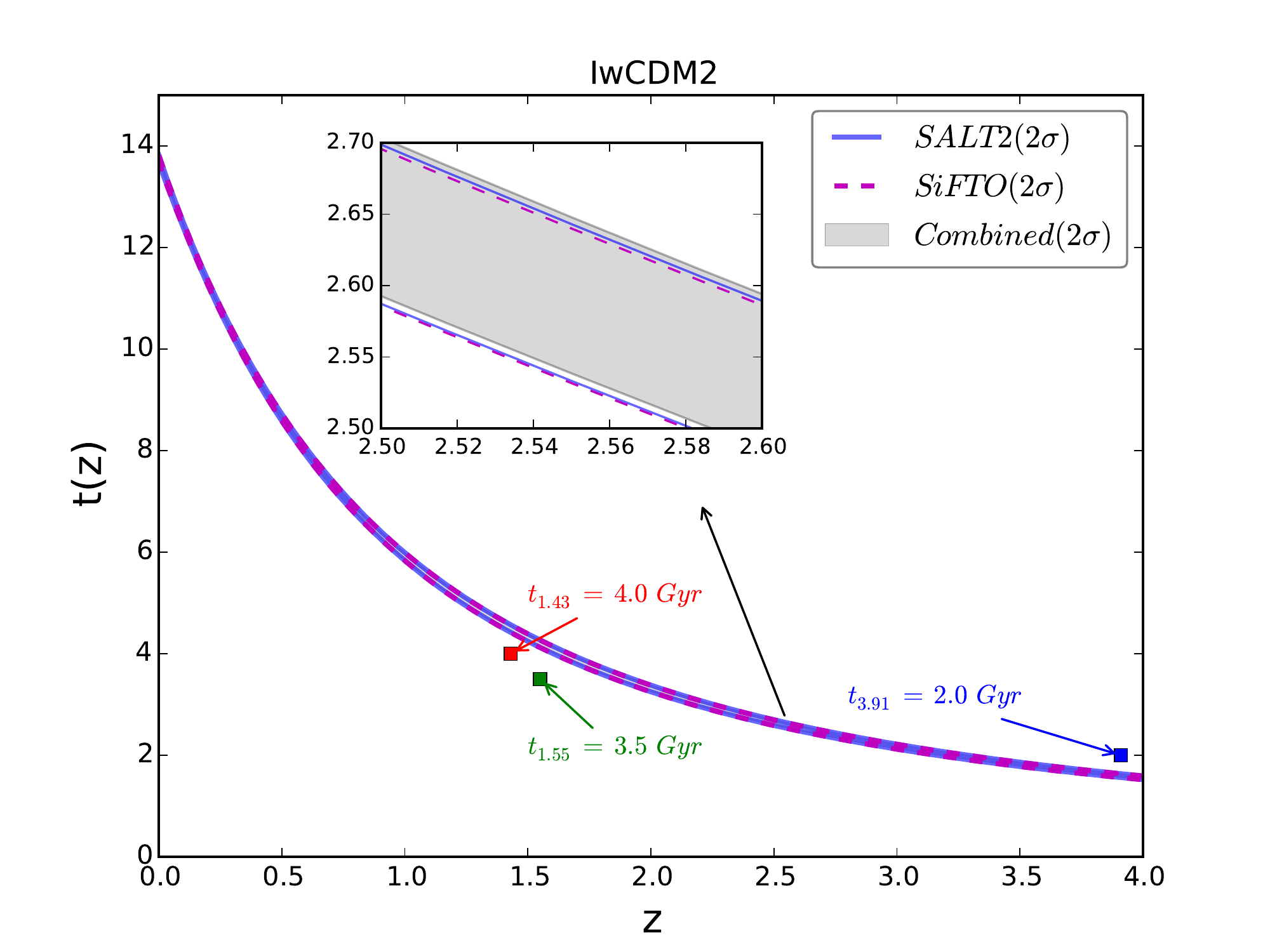}}
   \hspace{0.1\columnwidth}
  \resizebox{0.77\columnwidth}{!}{\includegraphics{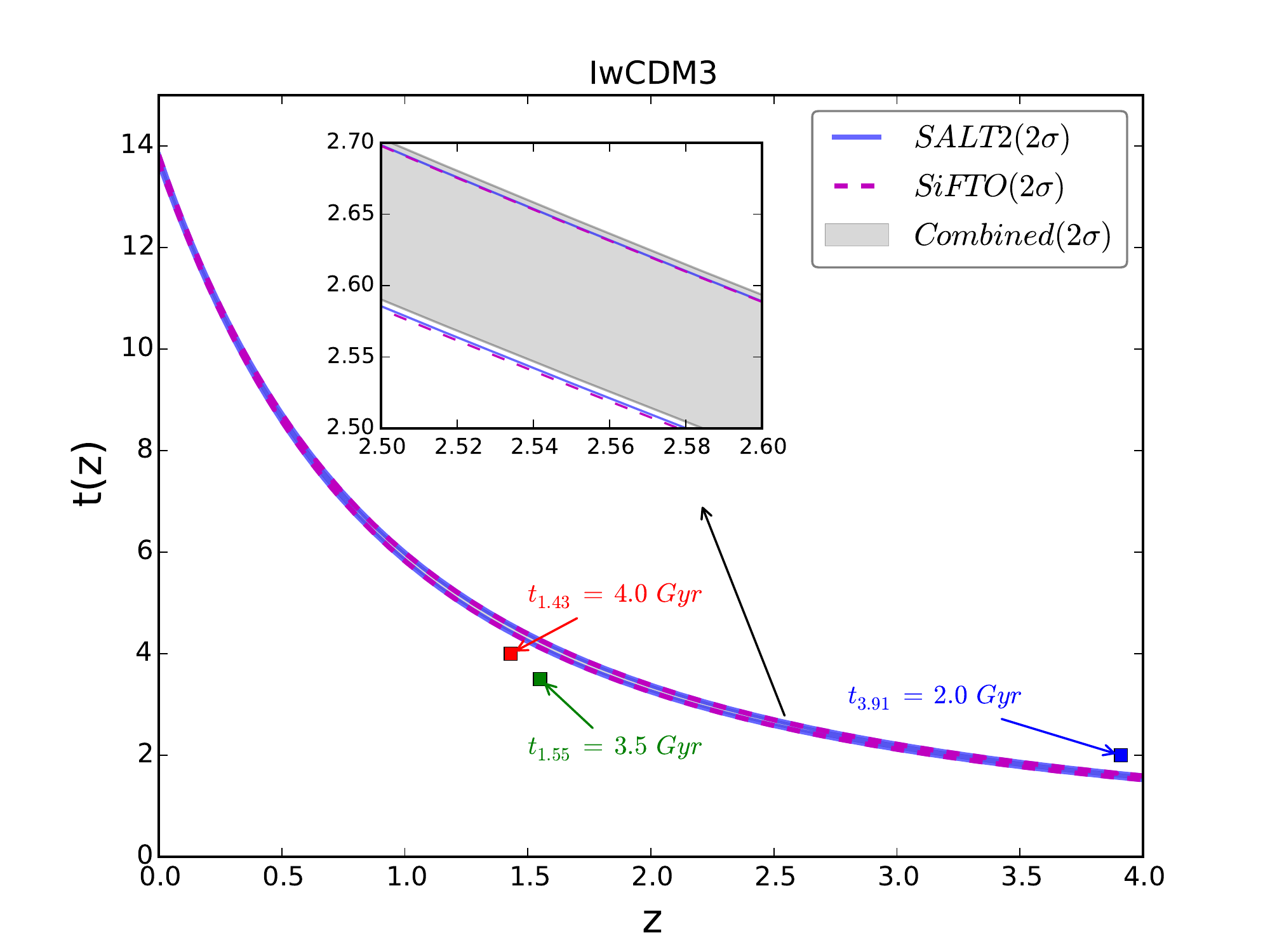}}
\caption{(color online). The 2$\sigma$ confidence regions of cosmic age $t(z)$ at redshift region $[0,4]$,
for the $w$CDM (upper left panel), the I$w$CDM1 (upper right panel), the I$w$CDM2 (lower left panel) and the I$w$CDM3 (lower right panel) model.
Three $t(z)$ data points, $t_{1.43}$, $t_{1.55}$ and $t_{3.91}$, are also marked by Squares for comparison.
``Combined'' (gray filled regions), ``SALT2'' (blue solid lines), and ``SiFTO'' (purple dashed lines) denote
the results given by the SN(Combined)+CMB+GC, the SN(SALT2)+CMB+GC, and the SN(SiFTO)+CMB+GC data, respectively.
}
\label{fig:tz}
\end{figure*}

The 2$\sigma$ confidence regions of cosmic age $t(z)$ at redshift region $[0,4]$ for the $w$CDM model and the three IDE models are plotted in Fig. \ref{fig:tz},
where the three $t(z)$ data points, $t_{1.43}$, $t_{1.55}$ and $t_{3.91}$,
are also marked by Squares for comparison.
We find that both $t_{1.43}$ and $t_{1.55}$ can be easily accommodated in all the four DE models,
but the position of $t_{3.91}$ is significantly higher than the 2$\sigma$ upper bounds of all the four DE models.
In other words, the existence of the old quasar APM 08279+5255 still can not be explained in the frame of IDE model.
This result is consistent with the conclusions of previous studies \cite{Alcaniz03,Wei07,WZ08,WZX08,WLL10,Yan14}.
In addition, the 2$\sigma$ regions of $t(z)$ given by different LCF are almost overlap,
showing that the impacts of different SNLS3 LCF can not be distinguished by using the age data of OHRO.

\begin{figure*}
  \centering
  \resizebox{0.77\columnwidth}{!}{\includegraphics{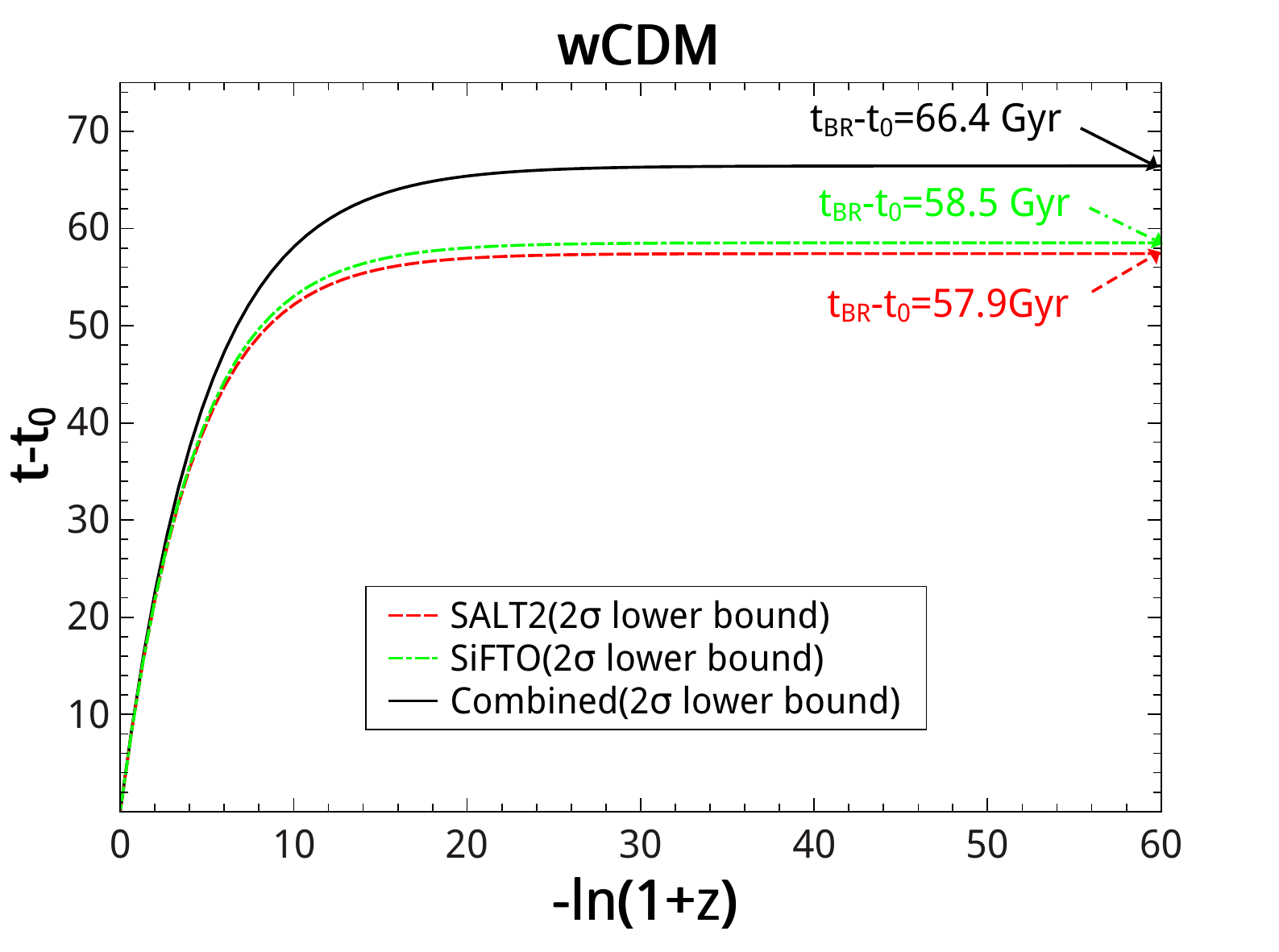}}
  \hspace{0.1\columnwidth}
  \resizebox{0.77\columnwidth}{!}{\includegraphics{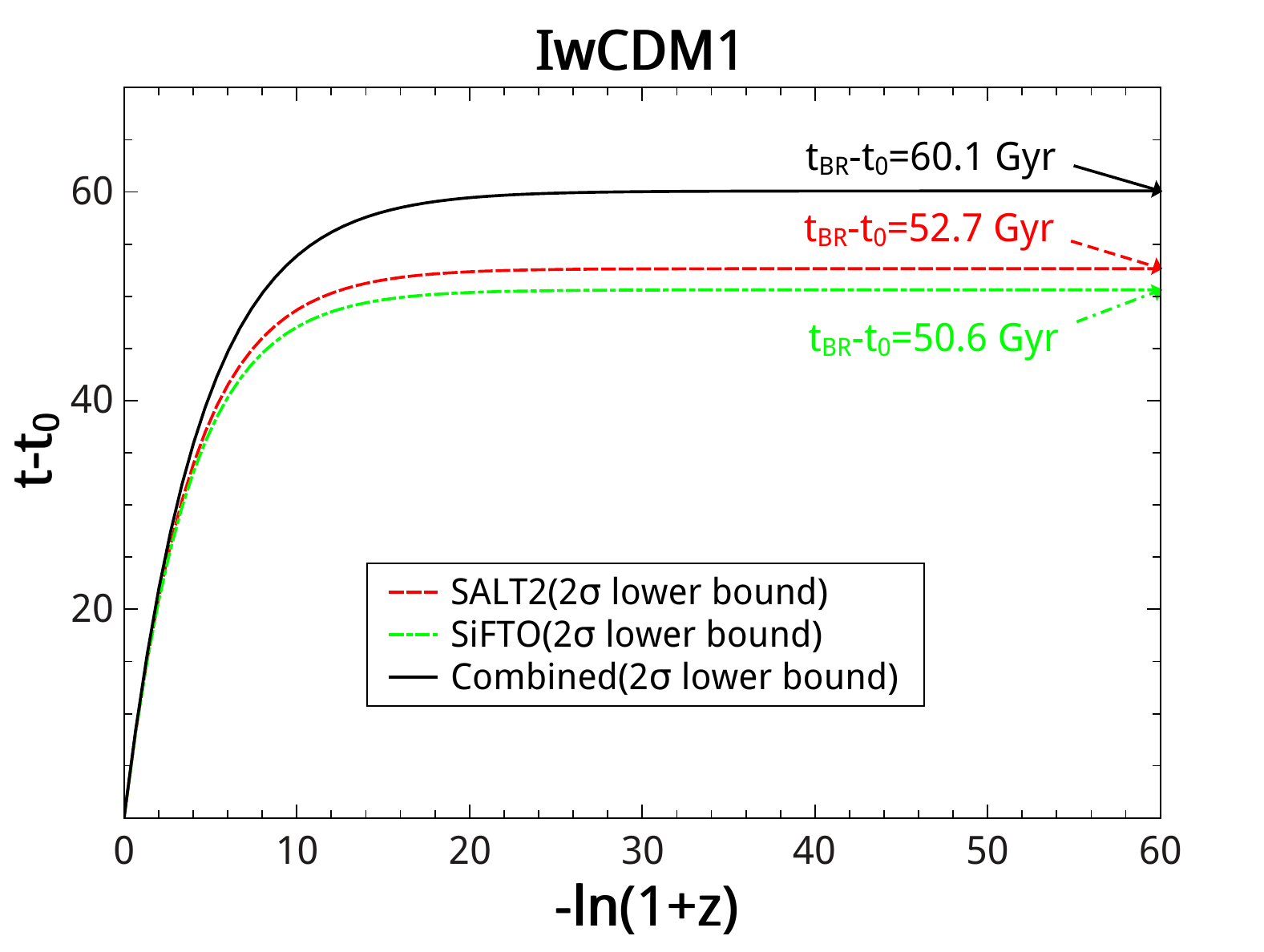}}
   \hspace{0.1\columnwidth}
  \resizebox{0.77\columnwidth}{!}{\includegraphics{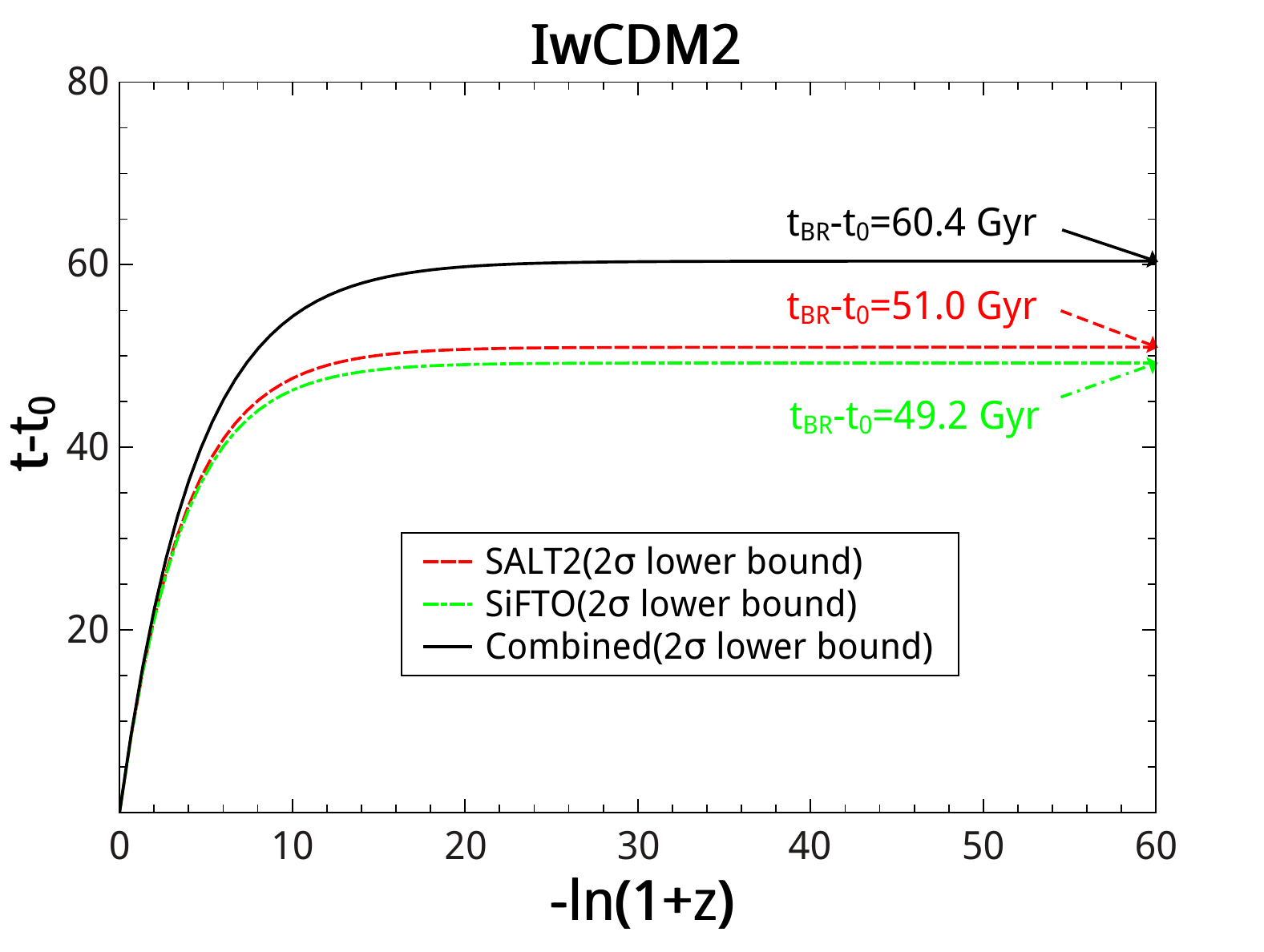}}
   \hspace{0.1\columnwidth}
  \resizebox{0.77\columnwidth}{!}{\includegraphics{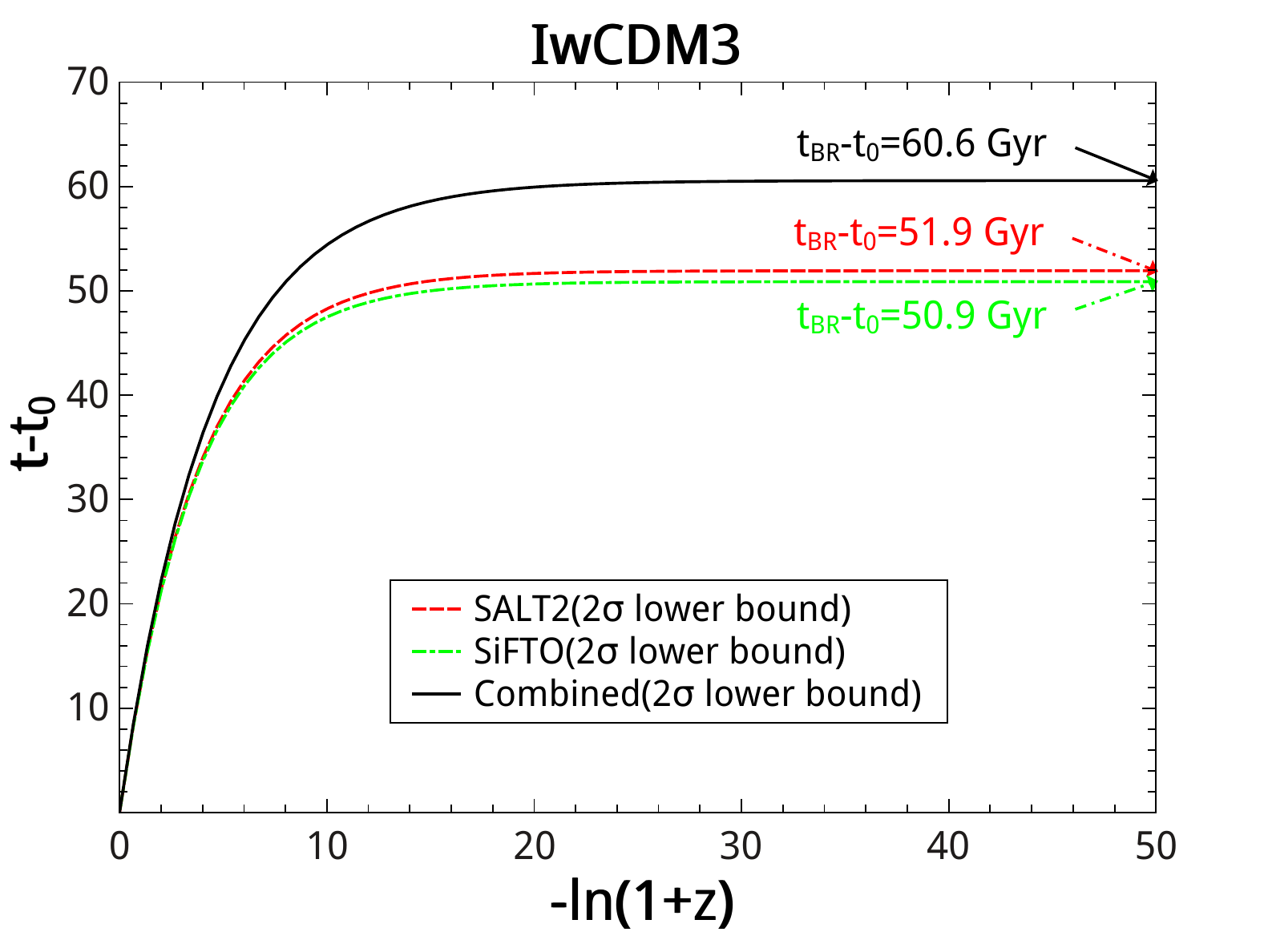}}
\caption{(color online). The 2$\sigma$ lower limits of the time interval $t-t_0$ between a future moment and today,
for the $w$CDM (upper left panel), the I$w$CDM1 (upper right panel), the I$w$CDM2 (lower left panel) and the I$w$CDM3 (lower right panel) model.
``Combined'' (black solid lines), ``SALT2'' (red dashed lines), and ``SiFTO'' (green dash-dotted lines) denote
the results given by the SN(Combined)+CMB+GC, the SN(SALT2)+CMB+GC, and the SN(SiFTO)+CMB+GC data, respectively.
The 2$\sigma$ lower bound values of $t_{BR}-t_0$ given by the three SNLS3 data are also listed on this figure.
}
\label{fig:fate}
\end{figure*}

We want to infer how far we are from a cosmic doomsday in the worst case.
So in Fig. \ref{fig:fate},
we plot the 2$\sigma$ lower limits of the time interval $t-t_0$ between a future moment and today, for the $w$CDM model and the three IDE models.
Moreover, the 2$\sigma$ lower limit values of $t_{BR}-t_0$ given by the three LCF are also marked on this figure.
\footnote{The 2$\sigma$ upper limits of $t_{BR}-t_0$ is infinite; in other words, the Universe will expand eternally}.
All the evolution curves of $t-t_0$ tend to the corresponding convergence values at $-ln(1+z) \simeq 20$.
The most important factor of determining $t_{BR}-t_0$ is the EoS $w$.
In a phantom dominated Universe, a smaller $w$ corresponds to a larger increasing rate of $\rho_{de}$;
this means that all the gravitationally bound structures will be torn apart in a shorter time,
and the Universe will encounter a cosmic doomsday in a shorter time, too.
Among the three SNLS3 LCF data,
the ``Combined'' sample always gives the largest $w$,
and thus gives the largest value of $t_{BR}-t_0$.
In addition, the values of $w$ given by the ``SALT2'' and the ``SIFTO'' sample are close to each other,
and thus the values of $t_{BR}-t_0$ given by these two samples are close to each other.

\section{Summary and Discussion}
\label{sec:conclusion}

As is well known, different LCF will yield different SN sample.
In 2011, based on three different LCF,
the SNLS3 group \cite{Conley11} released three kinds of SN samples, i.e., ``SALT2'', ``SiFTO'' and ``Combined''.
So far, only the SNLS3 ``Combined'' sample is studied extensively,
both the ``SALT2'' and the ``SiFTO'' data sets are seldom taken into account in the literature.
Moreover, the cosmological consequences given by these three SNLS3 LCF are not discussed before.
Therefore, the impacts of different SNLS3 LCF have not been studied in detail in the past.
The main aim of the present work is presenting a comprehensive and systematic investigation on the impacts of different SNLS3 LCF.

Since the interaction between different components widely exist in nature,
and the introduction of a interaction between DE and CDM can provide an intriguing mechanism
to solve the ``cosmic coincidence problem'' and alleviate the ``cosmic age problem'',
here we adopt the $w$CDM model with a dark sector interaction.
To ensure that our study is insensitive to a specific dark sector interaction,
three kinds of interaction terms are taken into account in this work.
In addition, to make a comparison, we also consider the case of $w$CDM model without dark sector interaction.

We have used the three SNLS3 data sets,
as well as the observational data from the CMB and the GC,
to constrain the parameter spaces of the $w$CDM model and the three IDE models.
According to the results of cosmology-fits,
we have plotted the cosmic evolutions of Hubble parameter $H(z)$, deceleration parameter $q(z)$,
statefinder hierarchy $S^{(1)}_3(z)$ and $S^{(1)}_4(z)$,
and have checked whether or not these DE diagnoses can distinguish the differences among the results of different LCF.
Furthermore, we have performed high-redshift cosmic age test using three OHRO, and have explored the fate of the Universe.

We find that for the $w$CDM and all the three IDE models:
(1) the ``Combined'' sample always gives the largest $w$; in addition, the values of $w$ given by the ``SALT2'' and the ``SIFTO'' sample are close to each other. (see table \ref{tab:combined_results} and Fig. \ref{fig:gamma_w});
(2) the effects of different SNLS3 LCF on other parameters are negligible (see table \ref{tab:combined_results}).
Besides, we find that the $\Lambda$CDM model is inconsistent with the three SNLS3 samples at 1$\sigma$ CL,
but is still consistent with the observational data at 2$\sigma$ CL.

Moreover, we find that the impacts of different SNLS3 LCF are rather small
and can not be distinguished by using $H(z)$ (see Fig. \ref{fig:hz}), $q(z)$(see Fig. \ref{fig:qz}), $S^{(1)}_3(z)$,
(see Fig. \ref{fig:statefinder3}), $S^{(1)}_4(z)$ (see Fig.\ref{fig:statefinder4}), and $t(z)$ diagram (see Fig. \ref{fig:tz}).
This result is quite different from the case of ``MLCS2k2'' \cite{Jha07} and ``SALT2'' \cite{Guy07},
where using ``MLCS2k2'' and ``SALT2'' LCF
will give completely different cosmological constraints for various models \cite{Bengochea11,Bengochea14}.
In addition, we infer how far we are from a cosmic doomsday in the worst case,
and find that the ``Combined'' sample always gives the largest 2$\sigma$ lower limit of $t_{BR}-t_0$,
while the results given by the ``SALT2'' and the ``SiFTO'' sample are close to each other (see Fig. \ref{fig:fate}).

Since the conclusions listed above hold true for  all the three IDE models,
we can conclude that the impacts of different SNLS3 LCF are insensitive to the specific forms of dark sector interaction.
In addition, these conclusions also come into existence for the case of the $w$CDM model.
Our method can be used to distinguish the differences among various cosmological observations
(e.g., see \cite{HLLW2015a,HLLW2015b}).

For simplicity, in the present work we only adopt a constant $w$, and do not consider the possible evolution of $w$.
In the literature, the dynamical evolution of EoS are often explored
by assuming a specific ansatz for $w(z)$ \cite{Chevallier01,Linder03,Gerke02,Wetterich04,Jassal05},
or by adopting a binned parametrization \cite{Huterer03,Huterer05,Huang09,WLL11,Li11,Gong13}.
To further study the impacts of various systematic uncertainties of SNe Ia on parameter estimation,
we will extend our investigation to the case of a time-varying $w$ in the future.

In a recent paper \cite{Betoule14}, based on the improved SALT2 LCF,
Betoule et al presented a latest SN data set (``joint light-curve analysis'' (JLA) data set), which consists of 740 SNe Ia.
Adopting a constant $\alpha$ and a constant $\beta$,
Betoule et al. found $\Omega_{m0}=0.295\pm0.034$ for a flat $\Lambda$CDM model;
this result is different from the result of SNLS3 data, but is consistent with the results of Planck \cite{Ade14}.
It would be interesting to apply our method to compare the differences between the SNLS3 and the JLA sample.
These issues will be studied in future works.

\section*{Acknowledgments}

We are grateful to the Referee for the valuable suggestions.
We also thank Prof. Yun Wang, Prof. Anze Slosar, Prof. Xin Zhang and Dr. Yun-He Li for helpful discussions.
SW is supported by the National Natural Science Foundation of China under Grant No. 11405024 and the Fundamental Research Funds for the Central Universities (Grant No. N130305007 and Grant No. 16lgpy50).
ML is supported by the National Natural Science Foundation of China (Grant No. 11275247, and Grant No. 11335012) and a 985 grant at Sun Yat-Sen University.

%----------------------------------------------------------------------
\bibliographystyle{aa}

\end{document}